\documentclass[apj]{emulateapj}
\usepackage{xspace}

\usepackage{hyperref}

\usepackage{epstopdf}
\usepackage{inputenc}
\usepackage[T1]{fontenc}
\usepackage{lmodern}
\usepackage{inputenc}
\usepackage{epsfig}
\usepackage{epstopdf}
\usepackage{amsmath}
\usepackage{lmodern}
\usepackage[T1]{fontenc}
\usepackage{natbib}
\usepackage{relsize}

\newcommand{\kms}{km~s$^{-1}$ }
\newcommand{\kmsp}{km~s$^{ -1}$}
\newcommand{\oi}{O~{\sc I}~}

\newcommand{\cii}{C~{\sc II}~}
\newcommand{\siii}{Si~{\sc II}~}
\newcommand{\siiii}{Si~{\sc III}~}

\newcommand{\vcen}{v$_{\text{cen}}$~}
\newcommand{\vn}{v$_{90}$~}
\newcommand{\oip}{O~{\sc I}}
\newcommand{\nvp}{N~{\sc V}}
\newcommand{\ciip}{C~{\sc II}}
\newcommand{\siiip}{Si~{\sc II}}

\newcommand{\siiiip}{Si~{\sc III}}
\newcommand{\feii}{Fe~{\sc II} }
\newcommand{\mgii}{Mg~{\sc II} }

\newcommand{\siivp}{S~{\sc IV}}
\newcommand{\vcenp}{v$_{\text{cen}}$}
\newcommand{\vnp}{v$_{90}$}
\newcommand{\wise}{{\it WISE} }
\newcommand{\wisep}{{\it WISE}}
\newcommand{\galex}{{\it GALEX} }
\newcommand{\galexp}{{\it GALEX}}
\newcommand{\sfrsd}{$\Sigma_{\text{SFR}}$ }
\newcommand{\sfrsdp}{$\Sigma_{\text{SFR}}$}

\newcommand{\mout}{$\dot{M}_{\text{o}}$ }
\newcommand{\mstar}{$M_\ast$ }
\newcommand{\mstarp}{$M_\ast$}
\newcommand{\vout}{$\text{v}_{\text{out}}$ }
\newcommand{\voutp}{$\text{v}_{\text{out}}$}
\newcommand{\moutp}{$\dot{M}_{\text{o}}$}
\newcommand{\ns}{$N_{\text{Si{\sc~II}}}~$}
\newcommand{\nsp}{$N_{\text{Si{\sc~II}}}$}

\newcommand{\vcircp}{v$_{\text{circ}}$}
\newcommand{\vcirc}{v$_{\text{circ}}$ }
\newcommand{\sfr}{M$_\odot$~yr$^{-1}$ }
\newcommand{\sfrp}{M$_\odot$~yr$^{-1}$}

\newcommand{\vesc}{v$_{\text{esc}}$ }
\newcommand{\vescp}{v$_{\text{esc}}$}

\shorttitle{Warm Outflow Scaling Relations}
\shortauthors{Chisholm et al.}

\begin{document}

\title{Scaling Relations Between Warm Galactic Outflows and Their Host Galaxies}
\author{John Chisholm\altaffilmark{1}}
\email{chisholm@astro.wisc.edu}

\author{Christina A. Tremonti\altaffilmark{1}}

\author{Claus Leitherer\altaffilmark{2}}

\author{Yanmei Chen\altaffilmark{3}}

\author{Aida Wofford\altaffilmark{4}}

\author{Britt Lundgren\altaffilmark{1}}

\altaffiltext{1}{Astronomy Department, University of Wisconsin, Madison, 475
  N. Charter St., WI 53711, USA}
\altaffiltext{2}{Space Telescope Science Institute, 3700 San Martin Drive, Baltimore, MD 21218, USA}
\altaffiltext{3}{Department of Astronomy, Nanjing University, Nanjing 210093, China}
\altaffiltext{4}{UPMC-CNRS, UMR7095, Institut d\rq{}Astrophysique de Paris, F-75014 Paris, France}

\begin{abstract}

We report on a sample of 51 nearby, star-forming galaxies observed with the Cosmic Origin Spectrograph on the Hubble Space Telescope. We calculate Si~{\sc II} kinematics and densities arising from warm gas entrained in galactic outflows. We use multi-wavelength ancillary data to estimate stellar masses (M$_\ast$), star-formation rates (SFR), and morphologies. We derive significant correlations between outflow velocity and  SFR$^{\sim 0.1}$, M$_\ast^{\sim 0.1}$ and v$_\text{circ}^{\sim 1/2}$. Some mergers drive outflows faster than these relations prescribe, launching the outflow faster than the escape velocity. Calculations of the mass outflow rate reveal strong scaling with SFR$^{\sim 1/2}$ and M$_\ast^{\sim 1/2}$. Additionally, mass-loading efficiency factors (mass outflow rate divided by SFR) scale approximately as M$_\ast^{-1/2}$. Both the outflow velocity and mass-loading scaling suggest that these outflows are powered by supernovae, with only 0.7\% of the total supernovae energy converted into the kinetic energy of the warm outflow. Galaxies lose {\it some} gas if log(M$_\ast$/M$_\odot$)~<~$9.5$, while more massive galaxies retain all of their gas, unless they undergo a merger. This threshold for gas loss can explain the observed shape of the mass-metallicity relation.

\end{abstract}


\section{INTRODUCTION}
\label{intro}

Stellar feedback shapes the evolution of galaxies. Complex interactions of supernovae, stellar winds and radiation pressure impart energy and momentum into surrounding gas. This deposition creates the multiphase structure of the interstellar medium (ISM) \citep{mckee, governato07, hopkins12a}, the discrepancy between the dark matter halo mass function and the baryonic mass function \citep{larson74, keres09, moster10}, the mass-metallicity relationship, and the enrichment of the intergalactic medium \citep{tremonti04, erb06, finlator08, andrews13}. By heating the densest gas in the ISM, the star-formation efficiency is reduced \citep{leroy08, bigiel08}, delaying and regulating the buildup of stellar mass \citep{katz96, springel03, hopkins12a}. 

Galactic outflows are the most noticeable result of stellar feedback, and they have been ubiquitously observed in star-forming galaxies across cosmic time \citep{heckman2000, weiner, chen10, martin12, rubin13, arribas2014}. Over the past decade, a coherent schematic of galactic outflows has been sketched: high mass stars deposit energy and momentum into the ISM, which thermalizes a large fraction of the gas into an over-pressurized hot (T > 10$^7$~K) plasma that accelerates out of the disk, entraining cool gas as it moves outward. Through this process, galactic outflows transfer nearly as much mass as is converted into stars \citep{rupke2005b, weiner, steidel10, newman12, rubin13}, at velocities comparable to the escape velocity \citep{heckman2000, rupke2005b, martin2005} to distances roughly 5-50~kpc from the starburst \citep{grimes2005, rubin11}. This significant mass loss regulates the star-formation within a galaxy, and produces many of the relations mentioned above. 

Theoretical models of galaxy evolution require galactic outflows to regulate the buildup of stellar mass. Simulations without feedback form stars too early, and too rapidly to match observations \citep{white91, oppenheimer06, piontek, hopkins12b}. However, properly accounting for the small mass and size scales associated with feedback makes the simulations too computationally expensive. It has become common for simulations to implement \lq\lq{}scaling relations\rq\rq{} to include uncertain, unresolvable, micro-scale physics. They relate the ejected gas to properties of the host galaxy to produce realistic galaxies \citep{springel03, oppenheimer06, vogelsberger}. These simulations include analytical relations of \lq\lq{}momentum-driven\rq\rq{} outflows, which parameterize momentum deposition from radiation pressure on dust grains. The momentum driven outflows scale the outflow velocity (\voutp) and mass loading efficiency (mass outflow rate divided by the star-formation rate; $\eta = $\moutp/SFR) with the circular velocity of the galaxy (\vcircp), such that \vout $\propto$ v$_{\text{circ}}$ \citep{springel03, murray05, oppenheimer06} and $\eta \propto \text{v}_{\text{circ}}^{-1}$ \citep{murray05}. Other possible driving mechanisms have been postulated, including thermal adiabatic expansion of supernova-heated gas, which produces scaling relations with \vout $\propto$ SFR$^{1/5}$ \citep{castor, weaver, ferrara00} and $\eta \propto \text{v}_{\text{circ}}^{-2}$ \citep{murray05, dutton, creasey}. In reality, the driving mechanism is probably a combination of the two mechanisms, each important during different times \citep{hopkins12b, vogelsberger13}. High resolution simulations of individual galaxies are nearly able to resolve the important physical scales to drive outflows. \citet{hopkins12b} produces four simulated galaxies, using both radiation pressure and thermal pressure, and find both mechanisms are essential to produce realistic galaxies. The simulations find $\eta$ of the simulated galaxies to scale as $\eta \propto \text{v}_{\text{circ}}^{-(0.9 \rightarrow 1.7)}$, with contributions from the two mechanisms dependent on the physical conditions within the galaxies. With theory honing in on the relevant physics, it is essential to check that the proposed scalings match observations. 

Observationally, galactic outflow scaling-relations have been difficult to study for three main reasons: 1) lack of suitable outflow tracers, 2) lack of sufficient host property dynamic range, and 3) lack of high quality data. The lack of suitable tracers has impacted the relations the most. In the optical, the most commonly used tracer is the Na~{\sc I} doublet (Na~D), near 5890~\AA. With an ionization potential of just ~5~eV, Na~D is easily ionized in the hard ultra-violet (UV) photon fields of starburst galaxies. This has prompted optical studies to use dusty, and therefore massive, starbursts \citep{martin2005, rupke2005b, chen10}. More recent studies use UV transitions, like Mg~{\sc II} and Fe~{\sc II}, that are redshifted into the optical \citep{weiner, erb12, martin12, kornei12, rubin13, bordoloi}. While these species survive the harsh UV radiation fields, by necessity the samples also probe brighter, more massive galaxies because they must be observed at higher redshifts (z > 0.35). The dynamic range is limited because only high redshift galaxies are sampled, making high quality observations more challenging. Often times averaged -- or \lq\lq{}stacked\rq\rq{} -- spectra are used to produce adequate signal-to-noise \citep{weiner, erb12, rubin13}, which may impact the derived results in an unknown manner. 

While measuring the trends is difficult, progress has been made; unfortunately, a clear picture has not emerged. \citet{martin2005} study the scaling relations with Na~D, and find an upper envelope of the centroid velocity (\vcenp) to scale with the star formation rate (SFR) as \vcen~$\propto SFR^{0.35}$. \citet{rupke2005b} expand the dynamic range of the mass and SFR to 4 orders of magnitude, and find \vout to scale strongly with the SFR and the circular velocity of the galaxy, with \vcirc scaling as \vout $\propto$ v$_{\text{circ}}^{0.85}$. Meanwhile, at higher redshifts studies using Fe~{\sc II} and Mg~{\sc II} find positive correlations with \vout and SFR \citep{weiner}, \mstar \citep{weiner, martin12, rubin13}, and star formation rate surface density ($\Sigma_{SFR}$; \citep{kornei12}). The uncertain trends predominately arise from how difficult it is to observe them. These issues can be mitigated by studying galaxies in the local universe -- where there is a large dynamic range -- with a tracer that robustly samples the dominant temperature regime of the outflow.

Here, we report on a Hubble Space Telescope (HST) Archival Study (Project ID: 13239), designed to use the wealth of the Cosmic Origin Spectrograph (COS) Archive to form a sample of 51 nearby (z < 0.27), star-forming galaxies (\autoref{sample}). In the rest frame UV, we use four \siii transitions, which probe the warm neutral, and partially ionized, gas between 6~$\times$~10$^3$ and 3~$\times 10^4$~K in galactic outflows \citep{mazzotta, draine}. From the \siii transitions, we derive kinematic and density estimates for the outflows (\autoref{data}), and combine these measurements with ancillary data from the Wide-field Infrared Survey Explorer (\wisep), GALaxy Evolution EXplorer (\galexp) and the Sloan Digital Sky Survey (SDSS; \autoref{masssfr}) to study outflow trends over 4 orders of magnitude in SFR and \mstar (\autoref{results}).  We finish by discussing the implications of these trends for galactic evolution (\autoref{discussion}).

In this paper we use $\Omega_M = .28$, $\Omega_\Lambda = .72$ and H$_0$ = 70~\kms~Mpc$^{-1}$ \citep{wmap}.

\section{SAMPLE AND DATA}
\subsection{Sample}
\label{sample}
\begin{figure*}[t]
\includegraphics[width = 1.0\textwidth]{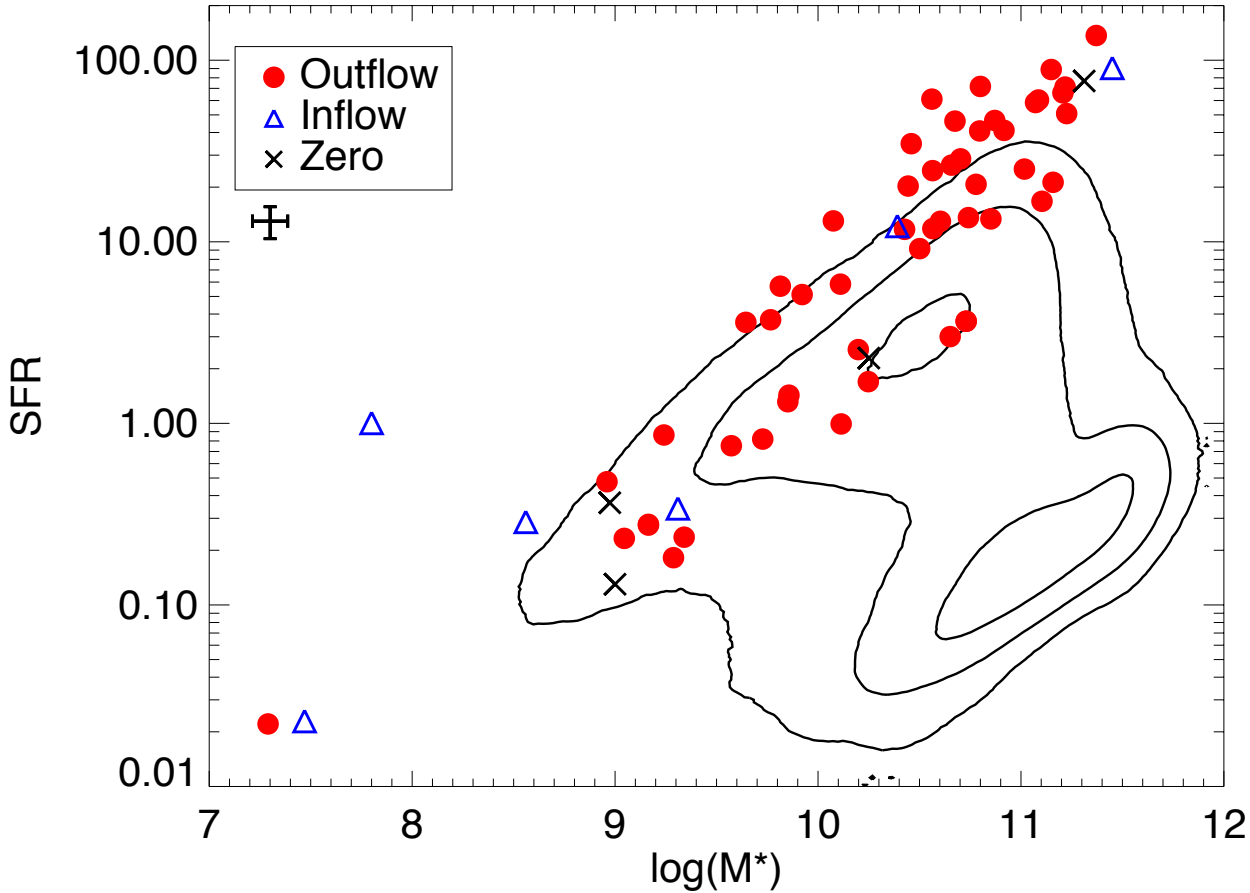}
\caption{Plot of the calculated SFR and stellar mass of the COS sample (see \autoref{masssfr}). Filled red circles are outflows at the 1$\sigma$ significance, blue triangles are inflows at the 1$\sigma$ significance, and the black xs are galaxies with absorption that is consistent with zero-velocity, as classified in \autoref{data}. A representative error bar is given below the legend. The contours enclose 25, 68 and 85\% of the full SDSS sample using the JHU-MPA calculation of stellar masses and SFRs \citep{kauffmann2003, brinchmann2004}. The sample covers both normal star-forming galaxies that lie on the SDSS main-sequence, and starburst galaxies that lie roughly 1~dex off the main-sequence.}
\label{fig:masssfr}
\end{figure*}

\begin{deluxetable*}{cccc}
\tablewidth{0pt}
\tablecaption{Previous Proposals Used}
\tablehead{
\colhead{Proposal ID} &
\colhead{PI} &
\colhead{References}  &
\colhead{Number} 
}
\startdata
11522 & J. Green   & \citet{france2010, wofford2013} & 6 \\
& & \citet{ostlin} & \\ 
11579 &  A. Aloisi & \citet{james} & 6 \\
11727  & T. Heckman & \citet{heckman2011} & 7 \\
12027 &  J. Green &  \citet{wofford2013} & 5 \\
12173 &   C. Leitherer &  \citet{claus2012} & 4 \\
12533 &  C. Martin & - & 2 \\
12583  &  M. Hayes & \citet{hayes, ostlin}  & 6 \\
12604  & J. Fox & \citet{fox2013, richter2013, fox14} &  1 \\
12928  &  A. Henry & \citet{jaskot}  & 5 \\
13017 &  T. Heckman & - & 9\\

\enddata
\tablecomments{Table of the 10 COS-GO/GTO proposals used to form the sample. Original proposal ID numbers and PIs are given in the first two columns, with associated references in the third column. The number of galaxies comprising the final sample of 51 is shown in the last column. The list of galaxies used are tabulated in \autoref{tables}}
\label{tab:sample}
\end{deluxetable*}

We create the sample by searching the MAST HST abstract archive for star-forming and starbursting galaxies observed with either the G130-M or the G160-M gratings on COS \citep{cos}. Care is taken to minimize Active Galactic Nuclei (AGN) contributions by using projects that study star-forming galaxies, as defined by the literature. We find a list of ten COS-GO/GTO proposals with 81 total galaxies, 14 of which do not make our signal-to-noise cut (\autoref{data}), for a initial sample of 62 galaxies. The proposals explore three broad categories: Ly$\alpha$ escape fractions measurements, direct metallicity measurements of star-forming galaxies, or characterization of the stellar populations of young stars. \autoref{tab:sample} lists the ten proposals comprising the sample, with the principle investigators (PIs) and related references. A full list of the sample, and the COS aperture locations are given in \autoref{tables}.

The sample probes a diverse range of galaxy types and spatial scales. COS has a 2.5\rq\rq{} diameter circular aperture which spans galaxy physical sizes from 0.05~kpc (for NGC 4449 at  z~=~0.0007) to 10~kpc (for GP0911+1831 at z~=~0.26). Therefore, some observations sample the entire galaxy, while other observations sample individual super-star clusters. Due to the diverse scientific goals of the various COS-GO/GTO projects, the stellar masses (\mstarp) and star-formation rates (SFR) also vary greatly. The SFRs range between 0.02~\sfr and 137~\sfrp, while the log(\mstarp/M$_\odot$) range from 7.29 to 11.4, four orders of magnitude for each parameter. \autoref{fig:masssfr} shows the physical parameters of the full sample (as calculated in \autoref{masssfr}), along with contours enclosing 25, 68 and 85\% of the Sloan Digital Sky-Survey (SDSS) DR7 galaxies. The sample extends from the SFR-\mstar relation, as denoted by the 68\% contours of the SDSS, up to 1~dex above the relation, probing both normal star-forming and starburst galaxies. The sample, and it\rq{}s derived properties, are tabulated in \autoref{tables}.

\subsection{COS Data}
\label{data}

\subsubsection{Data Reduction and Continuum Normalization}
The individual spectra are downloaded from the MAST server and processed through the CalCOS pipeline,  version 2.20.1. Individual exposures are combined using the methods outlined in Wakker et al. (in preparation), but here we give a brief description of the process. First, we calculate a wavelength-dependent velocity shift for each individual exposure by cross-correlating strong Milky Way and ISM absorption lines. 21-cm velocities from the LAB survey \citep{kalberla} help define zero-velocity for Milky Way absorption features, and are secondary references for velocity shifts. Individual exposures are shifted by the wavelength-dependent velocity offsets, and raw counts are combined and converted into flux values. Error measurements are calculated from the raw gross counts, after adjusting for the appropriate background levels (as defined in the COS handbook \citep{coshandbook}). Special attention is paid to edge effects and fixed pattern noise features, which are known to affect COS data \citep{danforth2010}. This method properly accounts for wavelength dependent fluctuations, while combining the flux and error arrays from the raw counts. 

The spectral resolution of the COS data depends on the filling of the circular aperture. For each spectrum we calculate an inverse variance-weighted, full-width at half maximum (FHWM) for six strong Milky Way lines (\siii $\lambda$1190, 1193, 1260, 1526, \siiii $\lambda$1206 and \cii $\lambda$1334, in the observed frame). We use this Milky Way FWHM as the spectral resolution. The spectral resolution of the sample ranges from 25~\kms to 215~\kmsp, with a median of 94~\kms ($R \sim 3200$). In one special case, M~83, the intrinsic absorption blends with the Milky Way absorption, therefore, we use the median FWHM of the sample for the FHWM for M~83. 

After the spectral resolution is measured, the spectra are deredshifted using redshifts from SDSS DR7 \citep{dr7}, or redshifts from the NASA/IPAC Extragalactic Database (NED)\footnote{The NASA/IPAC Extragalactic Database (NED) is operated by the Jet Propulsion Laboratory, California Institute of Technology, under contract with the National Aeronautics and Space Administration.}. To maximize signal-to-noise, each spectrum is binned by 10 pixels  (to 27~\kmsp), and smoothed by 3 pixels. We define four separate masks: Milky Way, geocoronal emission, strong stellar, and ISM (see \autoref{tab:mask}). Using all four masks, a power-law is fit between $1240-1400$~\AA, in the restframe, and the spectrum is normalized by the fit. Power-law slopes range from $-3.2-1.2$, with a median of -1.6. The Starburst99 theoretical models, described below, have power-law slopes between -2.5 and -3.5, indicating that the continuum is substantially reddened in many of the galaxies. After these steps, the spectra are in the rest-frame, and the intrinsic continuum shape is removed. We can now compare the spectra with fully theoretical models of the stellar continuum.

\begin{deluxetable*}{cccc}
\tablewidth{0pt}
\tablecaption{Masks Used}
\tablehead{
\colhead{Mask Name} &
\colhead{ Features Masked} & 
\colhead{ Mask Size}  &
\colhead{ Reference Frame} \\
\colhead{}&
\colhead{ } &
\colhead{(\kmsp)}  &
}
\startdata
Milky Way & \oip, \siiip, \siiiip, \siivp, \ciip & 400 & Observed   \\
Geocoronal & H~{\sc I}, \oip, N~{\sc I} &  750 & Observed   \\
ISM  & H~{\sc I}, \oip, \siiip, \siivp, \ciip & 600 & Rest\\
Stellar &   \siiip, \siiiip, \siivp, \ciip, & 300-650 & Rest \\
 &    C~{\sc III}, C~{\sc IV}, \nvp &  &  \\
\enddata
\tablecomments{Table of the four masks used in the data reduction. The masks are used at different stages to avoid contamination of the strong absorption and emission features. The masked features are listed, and the mask sizes are given in velocity units. The stellar masks take a range of values, depending on whether the line arises in a stellar wind (650~\kmsp) or a stellar photosphere (300~\kmsp). The origins of the lines are taken from \citet{leitherer11}.}
\label{tab:mask}
\end{deluxetable*}

The normalized spectra contain contributions from foreground Milky Way (in the observed frame),  stellar and ISM (in the rest-frame) absorption. To isolate the ISM absorption, we remove the stellar component using the high-resolution, fully theoretical Starburst99 model atmosphere libraries calculated using the WM-Basic code \citep{claus99, claus2010}. We use the fully theoretical models because the empirical models have lower spectral resolution than the COS spectra, and include interstellar absorption. The models require three inputs: a stellar metallicity, a star-formation law and a star-formation duration. Metallicities are derived from the stellar mass of the galaxy using the mass-metallicity relationship. Since the shape and slope of the mass-metallicity relationship strongly depends on the metallicity diagnostic used \citep{kewley08}, we consider two different mass-metallicity calibrations. The first relation,  from \citet{tremonti04}, uses the strong line method; while the second relation, from \citet{andrews13}, uses the direct method. The Starburst99 theoretical library has metallicities of 0.02, 0.2, 0.4, 1.0 and 2~Z$_\odot$ (where Z$_\odot$ = 0.02); and we choose the theoretical metallicity closest to the derived metallicities. This method does assume that the mass-metallicity relation holds true for all galaxies, an improper assumption for galaxies that fall below the mass-metallicity relation (Haro~11 for example). Galaxies with metallicities below the mass-metallicity relation will have older estimated ages than expected if we used the true metallicities, but we still fit the spectra with the models due to the age-metallicity degeneracy.

After the two possible metallicities are determined, a grid of Starburst99 models is created using an array of star-formation laws and burst ages. The models use a Kroupa IMF, with a power-law index of 1.3 for the low mass slope, an index of 2.3 for the high mass slope, a high mass cut off at 100~M$_\odot$, and the Geneva stellar evolution tracks for non-rotating stars, with high mass-loss. For consistency, the models are normalized with a power-law, using the same three non-stellar masks as the data. We then interpolate the models onto the same wavelength grid as the data, and convolve to the measured resolution of the data. Two star-formation laws are fit: instantaneous star-formation and constant star-formation. We use burst ages of 20~Myr and 100~Myr for the constant star-formation laws, and a grid between 4~Myr and 22~Myr, in steps of 2~Myr, for the instantaneous star-formation model. We then preform a $\chi^2$ analysis on the full, two metallicity grid to determine the best-fit Starburst99 stellar continuum model, and metallicity. The most commonly selected models have either a continuous star-formation law with an age of 100~Myr (38\% of the sample), or an instantaneous burst with an age of 14~Myr (38\% of the sample). The remaining models are instantaneous bursts with ages of 4~Myr (18\%), 8~Myr (2\%) and 22~Myr (2\%). We divide the data by the models to remove stellar absorption, which is minimal for \siiip, but significant for higher ionization species. This produces a spectrum that is largely a combination of ISM and Milky Way absorption features.

The outflow velocity is measured with respect to the stellar background within the COS aperture. The COS instrumental errors are expected to be on the order of $\pm$15~\kms \citep{coshandbook}, and larger errors are attributed to redshift errors, or galactic rotation. Unless the observations are made at the exact center of the galaxy (which most are), the stellar continuum will be offset in velocity due to the rotation of the stars in the potential of the galaxy. To account for these zero-velocity offsets, we cross correlate the best fit Starburst99 model with the data, using the geocoronal, Milky Way and ISM masks. The velocity offsets are normally distributed, with an average of 23~$\pm$~40\kmsp. We shift the spectra by the calculated offset, placing the spectra into the corrected rest-frames with respect to the stellar background.  Any remaining offset may arise from reported redshift errors.  After this correction has been made, the absorption features can be analyzed to characterize the outflows. 

\subsubsection{Velocity and Column Density Measurement}
\begin{figure*}[t]
\includegraphics[width = \textwidth]{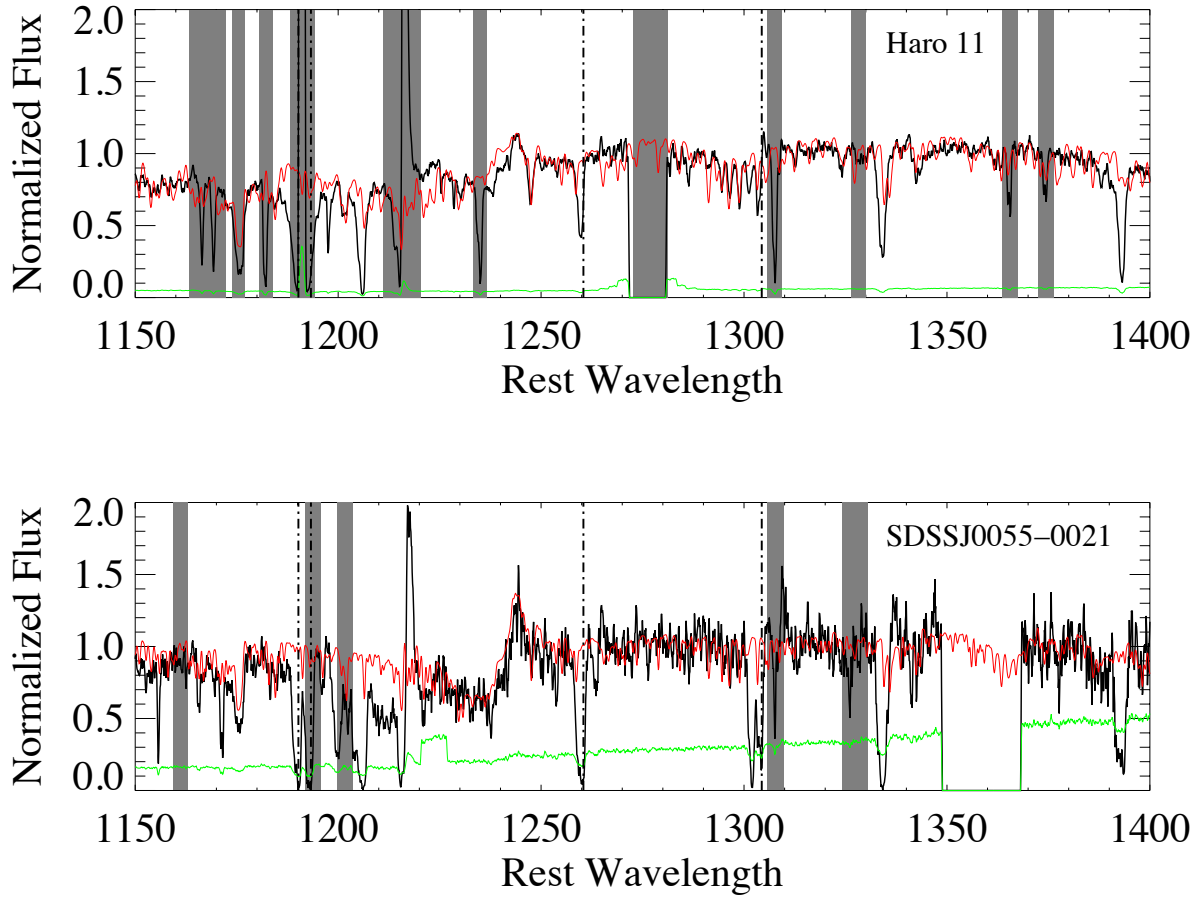}
\caption{COS spectra of Haro 11 (upper panel) and SDSSJ0055-0021 (lower panel) covering 1150-1400\AA\ in the rest-frame. The black line shows the normalized flux, and the green line shows the 1$\sigma$ error on the flux. The spectra contain many strong, broad ISM absorption lines, corresponding to different atomic species. The four resonant \siii features are denoted by vertical dot-dashed lines. Gray shaded areas mark possible Milky Way and geocoronal features. The red line displays the best fit Starburst99 stellar continuum model. The Haro 11 model has an instantaneous burst with an age of 14~Gyr, while SDSSJ0055-0021 has continuous star-formation with an age of 100~Myr. The stellar continuum model is used to remove the stellar contributions, and to correct for zero-velocity offsets.}
\label{fig:fullspec}
\end{figure*}

\autoref{fig:fullspec} shows two examples of rest-frame spectra between 1150 and 1400\AA, before the stellar continuum is removed. Broad (on average 2.7 times broader than the Milky Way absorption), blueshifted (on average offset by -113~\kms from the stellar continuum), ISM absorption lines are apparent throughout the spectra. We attribute the broad, blueshifted lines to gas entrained in galactic outflows \citep{heckman2000, claus2012}. To characterize the outflowing component we need a strong transition that probes the likely temperature and density ranges of the outflows. Simulations show that the expected temperature range of the bulk material of galactic outflows is expected to be between 10$^4$ and 10$^5$~K \citep{hopkins12b, creasey}. Therefore, we use the strong \siii transitions that probe the 10$^{3.8}- 10^{4.5}$~K temperature range \citep{mazzotta}. These \siii transitions also benefit from being both strong (see $f$-values in \autoref{tab:lines}) and cosmically abundant, efficiently probing the low density outflowing gas. 

To quantify the gas velocity and mass outflow rate, we fit four \siii resonant transitions with Gaussian profiles (see \autoref{tab:lines}). \siii 1260~\AA\ is the only line without a strong neighbor: \siii 1190 and 1193 form a doublet, with a  weak S~{\sc III} transition -52~\kms from the \siii 1190 line; while \siii 1304 has a strong \oi line separated by -504~\kmsp. We fit the doublet simultaneously, while the other two lines are fit individually to avoid biases from Ly-$\alpha$ absorption, Milky Way absorption and geocoronal emission which sometimes contaminate one or more of the lines. We simultaneously fit the nearby S~{\sc III} and \oi lines. We tie the S~{\sc III} transition to the \siii doublet, with the optical depth tied to the optical depth of \siii by the ratio of $f$-values and relative gas phase abundances in the Warm Ionized Medium \citep{jenkins}. The simultaneous fits for \oi do not have any tied parameters because the \oi line is more offset in velocity, and does not overlap with the \siii line as much as the S~{\sc III} line (the \oi line is 504~\kms from the \siii line, while the S~{\sc III} is 52~\kms from the \siii line).

\begin{deluxetable}{cccc}
\tablewidth{0pt}
\tabletypesize{\scriptsize}
\tablecaption{Lines Used}
\tablehead{
\colhead{ Wavelength} &
\colhead{ Nearby Lines} &
\colhead{ Velocity Separation} &
\colhead{ $f$-value} \\
\colhead{\AA} &
\colhead{ } &
\colhead{(km s$^{-1}$) } &
\colhead{}
}
\startdata
1190.42 & \siii 1193, S~{\sc III} 1190 & 773, -52 &  0.28   \\
1193.28 & \siii 1190 &  -773 & 0.58   \\
1260.42  & -  & -  & 1.22\\
1304.37 &  \oi 1302 & -504 & 0.09 \\
\enddata
\tablecomments{Table of the four \siii lines measured. The lines listed in the Nearby Lines column are fit simultaneously, while each row is fit individually. S~{\sc III} 1190 is a very weak ($f$-value of 0.02) line. \siii 1260 is the only line without a strong neighboring line. The S~{\sc III} line is fit with parameters tied to \siii 1190, while \oi has no tied parameters. With a velocity resolution of 94~\kmsp, the blending of the \oi transition and the \siii 1304~\AA\ line is minimal. $f$-values are taken from \citet{transition}, and differ by a factor of 14 between the strongest and weakest lines. Wavelengths are from \citet{wavelength}, and are downloaded from the NIST database \citep{nist}}
\label{tab:lines}
\end{deluxetable}

Before fitting the absorption lines, we renormalize the continuum with a 2$\sigma$ clipped mean of the continuum between $\pm$1500-2500~\kms from the line center, in the rest frame. This is necessary because there are large ISM contributions to the continuum that are not removed with the stellar continuum, especially broad Ly-$\alpha$ absorption near the \siii 1190 doublet. The lines are fit with a Gaussian profile using MPFIT, a robust non-linear least squares routine \citep{mpfit}, with the inverse-variance weighted Milky Way FWHM as the spectral resolution. As described by \citet{rupkee2005}, the profiles are parameterized with a centroid velocity (\vcenp), Doppler b-parameter (b), optical depth ($\tau$), and covering fraction (C$_f$). MPFIT returns estimates and errors for each parameter, and we set one-tenth the spectral resolution as the minimum velocity error.

\begin{figure*}
\includegraphics[width = 1.0\textwidth]{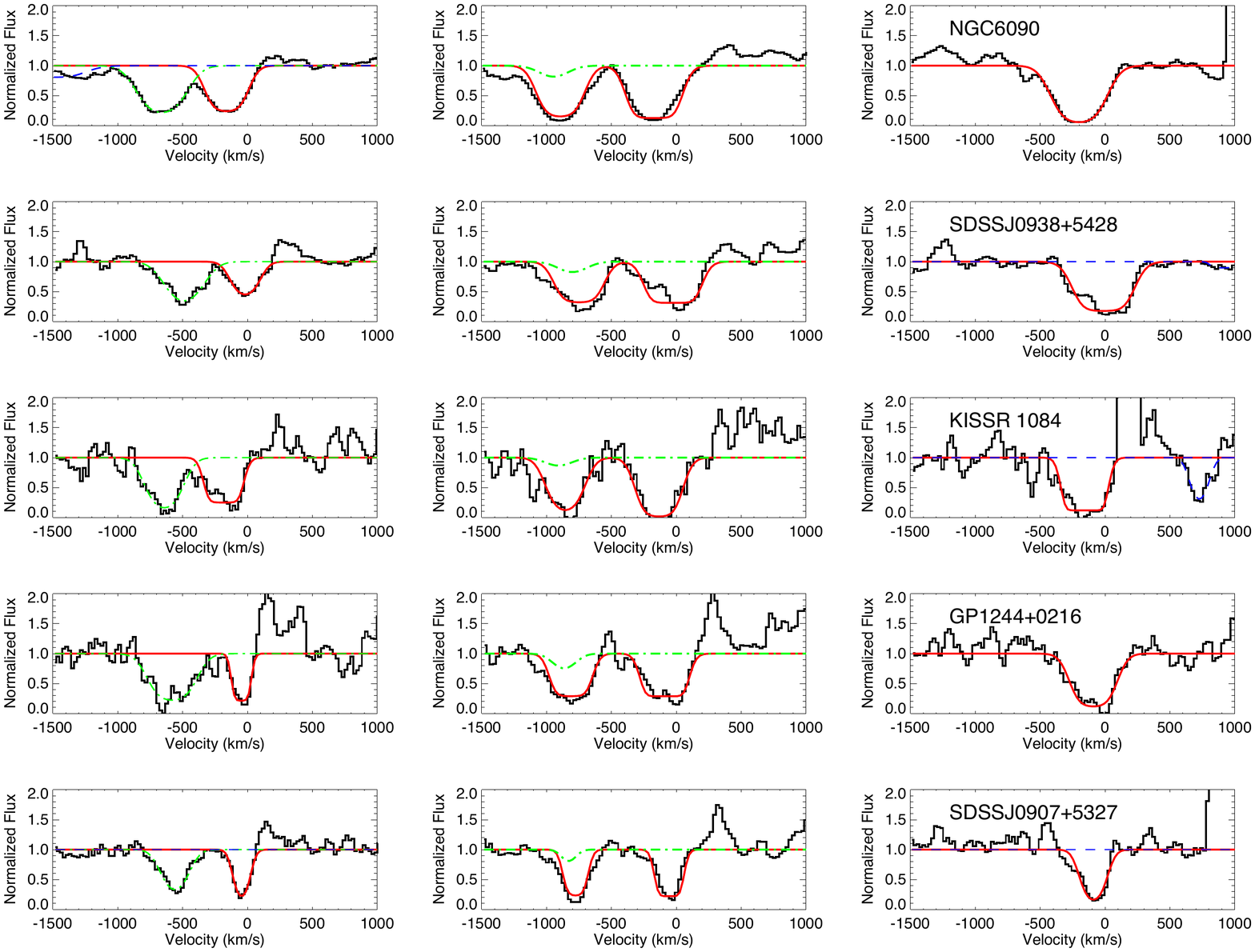}
\caption{Example fits of the four \siii transitions for five representative galaxies, with names given in the far right panel. From the left panel to the right panel, the transitions are \siii 1304, the \siii 1190 doublet, and \siii 1260. The f-values for the transitions increase from left to right (see \autoref{tab:lines}). The \oi 1302 and the S~{\sc III}~1190 line in the left and middle panels, respectively, are simultaneously fit, and shown as green dot-dashed lines. The simultaneous fits of \oi have no tied parameters, while S~{\sc III} has all the parameters tied to \siiip. Whenever Milky Way lines contaminate the absorption, the fit is shown as a blue dashed line (i.e. the \siii 1260 transition for KISSR~1084). }
\label{fig:fits}
\end{figure*}

\begin{figure}
\includegraphics[width = 0.5\textwidth]{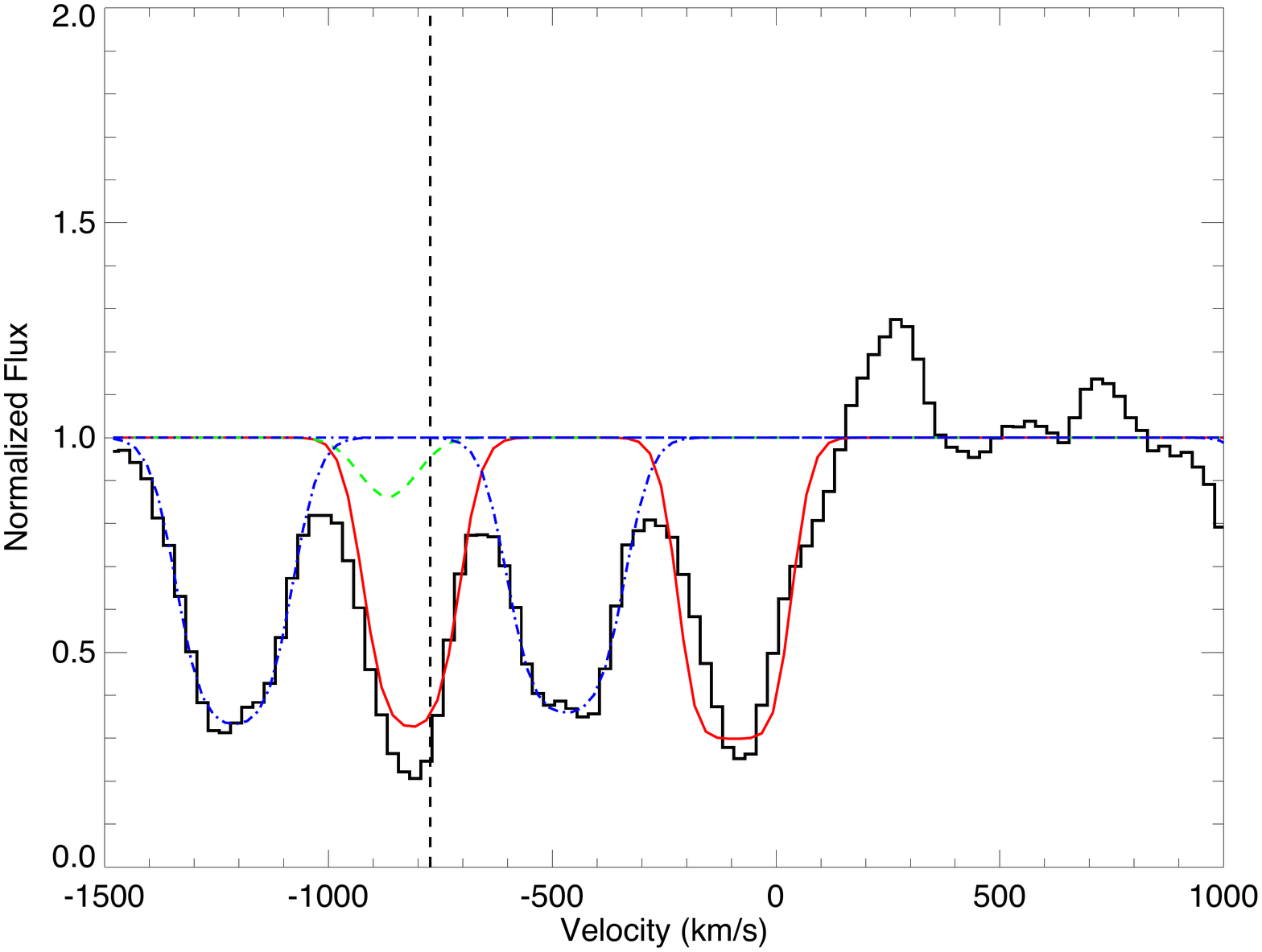}
\caption{Example fit (red solid line) of the \siii 1190 and 1193\AA\ doublet for NGC~5253 (data is the black histogram line). Heavy Milky Way foreground absorption contaminates the absorption profiles of the background galaxy. A simultaneous fit of the Milky Way absorption is shown as a blue dot-dashed line. The velocity of the Milky Way absorption is constrained within 100~\kms of the observed frame velocity. The S~{\sc III} profile fit is included as a green dashed line. The zero-velocity of the \siii 1190 line is shown as a vertical dashed line.}
\label{fig:mw}
\end{figure}

Example fits for five galaxies, representative of the typical signal-to-noise, are shown by solid red lines in \autoref{fig:fits}, with the simultaneous fits of nearby transitions shown as dot-dashed green lines. We simultaneously fit strong Milky Way absorption features with a velocity constrained to $\pm$100~\kms around the observed frame zero-velocity. \autoref{fig:mw} shows an example fit of the \siii 1190 doublet with heavy Milky Way contamination, where the profiles are successfully recovered, and velocities agree within 1$\sigma$ for all four transitions. 

While we use \vcen as a density-weighted average velocity of the outflowing gas, we also need a measure of the maximum outflow velocity. Previous studies have either used the velocity at a percent of the continuum level \citep{martin2005, weiner, erb12}, or a combination of \vcen and line width \citep{rupkee2005, arribas2014}. We choose to use the velocity at 90\% of the continuum (\vnp) of the best fit model to measure the maximum velocity. This choice is motivated by the fact that \vnp does not rely as strongly on the fitting process, and assumed velocity profile of the gas. The errors for \vn are calculated by bootstrapping \vn with 1000 simulations. For the sample, the median \vcen and \vn are -113 and -391~\kmsp, respectively. We measure the equivalent width of each line using the best-fit model, and bootstrap the errors with 1000 simulations. Finally, we estimate the column density (\nsp), at line center, using the relation from \citet{spitzer}
\begin{equation}
N_{\text{\siii}} = \frac{\tau_0 b} {1.497 \times 10^{-15} \lambda_0 f}
\end{equation}
Where the oscillator strengths ($f$) and wavelengths ($\lambda_0$) come from \citet{transition} and \citet{wavelength}, respectively (see \autoref{tab:lines}). The column densities for the sample range between 10$^{13.6}$ and 10$^{16.3}$~cm$^{-2}$, with a median of 10$^{14.8}$~cm$^{-2}$. The column density acts as a transition independent measure of the amount of material swept up in the outflow.

\begin{figure}[!h]
\includegraphics[width = 0.5\textwidth]{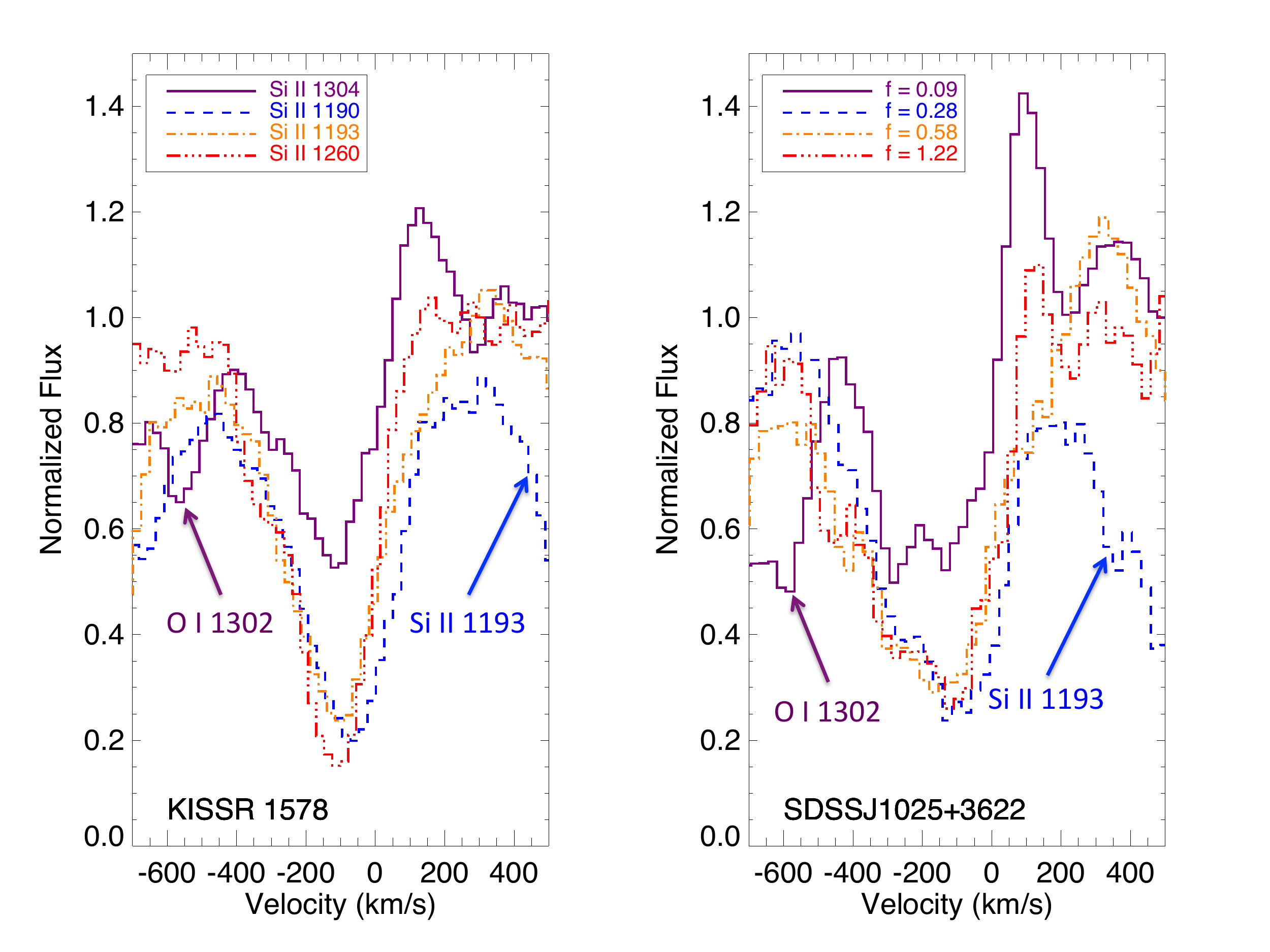}
\caption{The four \siii transitions for two galaxies: KISSR~1578 (left panel) and SDSSJ1025+3622 (right panel). The four different lines have varying $f$-values, as shown in the legend on the right. The profiles for the higher $f$-values are similar, while \siii 1304, the lowest $f$-value transition, has a markedly different profile. Various contaminating lines are illustrated and labeled by arrows. For the strongly saturated lines, the absorption is not black, indicative of a non-zero covering fraction. Hints of emission are seen in \siii 1304, indicative of a slight P-Cygni profile.}
\label{fig:profiles}
\end{figure}

\begin{figure}
\includegraphics[width = 0.5\textwidth]{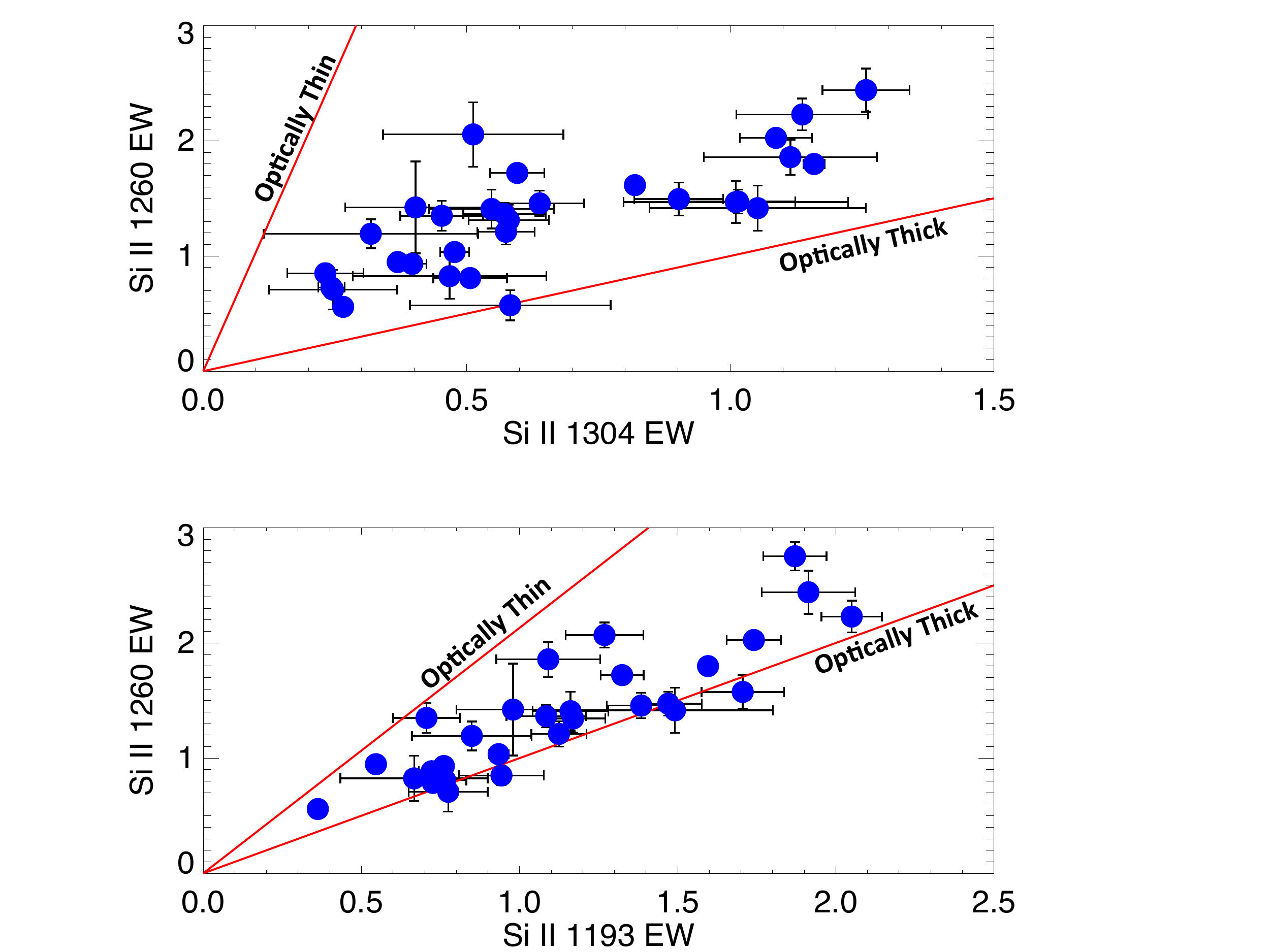}
\caption{Plot of the equivalent widths for two, low $f$-value \siii transitions (1304 and 1193) against the largest $f$-value \siii transition, 1260~\AA. The red lines marks the expected ratio in the optically thin and thick regimes. The 1304 transition (upper panel) has the lowest $f$-value, and points predominately cluster between the optically thin and thick regions. The 1193 transition (lower plot) predominately resides along the optically thick line, and suffers larger saturation effects. }
\label{fig:ewrat}
\end{figure}

With four separate transitions, we can explore how the measured quantities vary from transition to transition, and illustrate the important limitations and uses of each quantity. \autoref{fig:profiles} depicts the four \siii absorption lines at 1304, 1190, 1193, and 1260\AA, with $f$-value ratios of 1:3.1:6.4:13.5, respectively (\citep{transition}; see \autoref{tab:lines}). The large $f$-value transitions have similar profiles, indicative of saturation. Conversely, \siii 1304, the lowest $f$-value transition, has a distinctly different profile. \autoref{fig:ewrat} shows the ratio of the equivalent widths for the \siii 1304, 1260 and 1193 transitions, along with demarcations of the optically thin and thick regimes. While many of the ratios for \siii 1193 lie near the optically thick limit (lower line in the lower panel), the 1304 ratios predominately populate the transition region between optically thin and optically thick (upper panel). Therefore, we use the column density from the \siii 1304 transition, when possible, as a proxy of the \siii column density. When \siii 1304 is not available (due to geocoronal emission, chip gaps, etc.), the \siii 1193 column density is used; and if the 1193 is not available the 1260 measurement is used to measure the column density. While the 1304 lines are not as saturated, the features are much weaker, and noisier; an unavoidable trade-off for using the weaker features. Occasionally the 1304 measurement has signal-to-noise below 0.5, in these cases the higher signal-to-noise 1193 column density measurement is used instead. We eliminate one galaxy because there is not a column density measurement with signal-to-noise greater than 0.5.  In seven cases the 1304 line is not resolved (the line width is less than the Milky Way FWHM), in these cases we opt for the resolved, higher $f$-value transitions.

Since we are using different transitions, we first check that the \ns measurements from the different transitions are drawn from reasonably similar distributions. We preform a K-S test on the N$_{1260}$ and the N$_{1304}$ distributions, and find a p-value of 2.8~$\times$ 10$^{-6}$, indicating that the different transitions arise from different distributions. We find a strong (3$\sigma$) relationship between the column densities such that
\begin{equation}
N_{1304} = 5.3 \pm 1.2 \times N_{1260} + 2.0\times 10^{14} \text{~cm$^{-2}$}
\end{equation}
After the correction is made, a K-S test reveals that the column densities are drawn from statistically similar distributions, and we use these corrected values for \nsp, and the errors from the fit are propagated through to the \ns errors. This approach is similar to the saturation corrections made in \citet{savage91}, and accounts for the saturated, unresolved components.

\begin{figure}
\includegraphics[width = 0.45\textwidth]{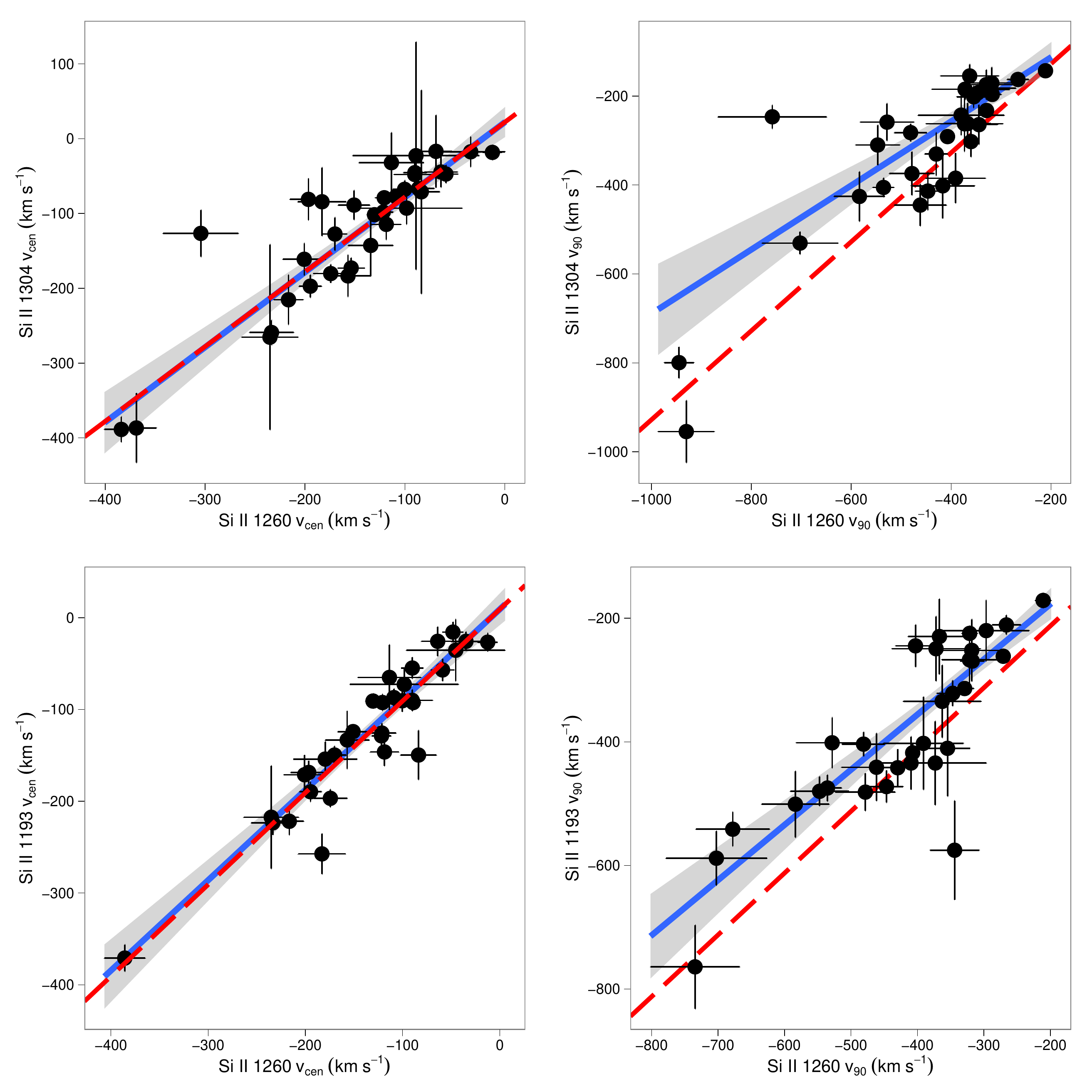}
\caption{Comparison of the velocity measurements for three \siii transitions. The top two panels show a comparison between the 1304 and 1260 transition, while the bottom two panels show the comparison between the 1193 and 1260 transition. The two panels on the left compare \vcenp, while the two panels on the right compare \vnp. Variance weighted regression lines are shown in blue, with the shaded region enclosing the 95\% confidence region. Red dashed lines show a linear relation in each panel. For both the 1193 and the 1304 transitions, \vcen shows a consistent relation with unity slope, while the \vn relations deviate from unity. The different \vn relations arise because the 1304 line has a lower f-value, and consequently only traces the highest density gas found at lower velocities. For this reason we use the inverse-variance weighted average of all the available lines for \vcen measurements, while we use the highest $f$-value transition for \vnp.}
\label{fig:velocities}
\end{figure}

Finally, in \autoref{fig:velocities}  we explore the relationship between the measured \vout for the four transitions. There is a linear (slope of one, intercept of 0) trend between the \vcen for all the transitions, while the \vn relationship departs from unity, as shown by the red line in \autoref{fig:velocities}. When multiple transitions are present, we use the inverse variance weighted mean velocity of all transitions to estimate \vcenp, while we always use the transition with the highest $f$-value to estimate \vnp. \citet{martin09} and \citet{grimes2009} find a similar effect with \mgii and \feii transitions, such that the covering fraction (or absorption strength) decreases with increasing velocity. Only the strongest transitions reliably probe the high velocity gas because the high velocity gas has the lowest covering fraction (or column density). In summary, when multiple transitions are available, we use the inverse-variance weighted \vcenp, and the \vn of the highest $f$-value transition to estimate \vcen and \vnp.

We classify outflows by requiring a variance weighted \vcen less than zero at the 1$\sigma$ significance level. Inflows are classified by a \vcen greater than zero at the 1$\sigma$ level, and zero velocity absorption can neither be classified as outflow, nor inflow, at the 1$\sigma$ significance level. After these classifications are made, 50 of the 61 galaxies are classified as having outflows. There are two observations of two different star clusters in M~83, and we use both. We derive \vcenp, \vnp, equivalent widths and column densities for these 51 observations. The measured values for each galaxy are included in \autoref{tab:outflow}. Below, we combine these measurements with \mstarp, SFR and \sfrsd to study trends in outflow properties with host galaxy properties.

\section{CALCULATION OF HOST PROPERTIES}
\label{masssfr}

We wish to study the trends of outflow properties (\vcenp, \vnp, and \nsp) with the properties of their host galaxies. To create a consistent data set, we calculate \mstar and SFR using the all-sky data from \galex and \wise. This reduces the systematics of joining multiple data sets, and allows for the consistent treatment of galactic properties.  In \autoref{masses} we discuss the calculation of \mstarp, in \autoref{sfr} we discuss the calculation of SFRs, in \autoref{sfrsd} we discuss our calculation of \sfrsdp, and in \autoref{incline} we characterize the inclinations and morphologies. All of the values derived here are tabulated in \autoref{tab:target} and \autoref{tab:gal}.

\subsection{Stellar Masses}
\label{masses}
In order to calculate the stellar mass (\mstarp), we use the all-sky infrared (IR) observations from \wise \citep{wise} to probe the Rayleigh-Jeans tail of the stellar blackbody emission. The IR is ideal for \mstar calculations because the IR probes the low mass stars that contribute the most to the total stellar mass, while having little attenuation from dust. \wise has four bands at 3.4 (W1), 4.6 (W2), 12 (W3), and 22 $\mu$m (W4). In general, W1 samples stellar emission, W2 samples both stellar emission and hot dust emission, W3 samples PAH absorption and dust emission, and W4 samples dust emission. Apparent magnitudes are downloaded from the NASA/IPAC Infrared Science Archive. The exponential fits are used for the 16 galaxies that are resolved by \wisep, otherwise the instrumental profile fits are used.  M~83 is resolved, but it has a \wise center offset from the 2MASS center, and the exponential fit is not calculated. Thus, we use the resolved \wise magnitudes from \citet{jarrett2013}. Apparent magnitudes are aperture corrected using the prescription from \citet{jarrett2011} and the \wise Explanatory Supplement \citep{wisesupp}.

The absolute magnitudes are calculated using the tabulated distances (see \autoref{tab:target}). For the galaxies within 40~Mpc, we use redshift independent distance from NED, while for the more distant galaxies we calculate distances using the luminosity distance and the redshift. K-corrections are calculated using equations 5.65 and 5.71 from \citet{longair}:
\begin{equation}
K\left( z, \alpha \right) = - 2.5 \text{log}_{10} \left[ \left( 1+z \right)^{1-\alpha} \right]
\label{eq:kcor}
\end{equation}
where $\alpha$ is the spectral index, defined as F$_\nu$ $\propto$ $\nu^{-\alpha}$. We calculate $\alpha$ from the \wise colors and a spline fit to the \wise color-index relation from Table 1 of \citet{wise}. K-corrections range between $-0.2$ and $+0.2$ magnitudes for W1, and $0.0$ and $-0.5$ magnitudes for W4. The W1 k-corrections change sign because the stellar continuum dominates redward of 4~$\mu$m, while the dust emission begins to dominate blueward of 4~$\mu$m. Since many of the galaxies have usually high dust temperatures and very young stellar populations \citep{engelbracht}, the location of the cross-over varies from galaxy to galaxy.   

Once the absolute magnitudes (M$_{W1}$) are K-corrected, the W1 luminosity is calculated using the relation from \citet{jarrett2013}:
\begin{equation}
L_{\lambda, W_1} \left(L_\odot \right) = 10^{-0.4 \left(M_{W1}-3.24 \right)}
\end{equation}
We calculate \mstar from L$_{\lambda, W_1}$ using a Chabrier IMF and the constant 3.6~$\mu$m mass-to-light ratio ($\Upsilon_{3.4}$) of $0.6 \pm 0.07$  \citep{meidt}. \citet{meidt} find that $\Upsilon_{3.6}$ accurately accounts for the mass of starbursts with a weak scatter due to the stellar population age. We do not use a color-dependent $\Upsilon_{3.6}$ because of the unusually warm dust temperatures, which redden the W1-W2 colors, and decrease many of the \mstar measurements. We find unacceptable disagreement (order of magnitude) between calculated \mstar and literature values for some galaxies in the sample when we use a color-dependent $\Upsilon_{3.6}$. We estimate \mstar uncertainties of 20\%, which includes the 11\% uncertainties on $\Upsilon_{3.6}$, the 5\% \wise magnitude uncertainties, and 15\% distance uncertainties.

\subsection{Star-formation Rates}
\label{sfr}

The star-formation rate (SFR) is an important indicator of the energy injected into the ISM. The most direct method to measure the SFR is the UV continuum from young, massive stars. However, dust severely attenuates the UV continuum, obscuring a large fraction of the star-formation. The sample spans a wide range in galaxy types, from dwarf star-forming galaxies to heavily dust-obscured ultra-luminous infrared galaxies (ULIRGS),  therefore, the SFR indicator must be sensitive to both obscured, and unobscured star-formation. We use a combination of the IR star-formation rate (SFR$_{\text{IR}}$; measured with \wise W4) and the UV star-formation rate (SFR$_{\text{UV}}$; measured by the FUV magnitudes with effective wavelength of 1516\AA\ from \galex) to measure the total star-formation rate (SFR$_{\text{Tot}}$).

To calculate SFR$_\text{IR}$ we use the 22~$\mu$m dust emission measured by the W4 band on \wisep.   W4 apparent magnitudes are aperture and k-corrected, as described above. Luminosities ($L_\nu$) are calculated using the color-corrected zero point of 2.16$\times$10$^{-56}$~L$_\odot$~Hz$^{-1}$ \citep{jarrett2011} and either the redshift independent distance, or the luminosity distance. The W4 luminosities are reduced by 8\% to account for the red calibration uncertainty in W4 \citep{jarrett2011, wisesupp, jarrett2013}. The SFR$_\text{IR}$ is calculated using the relation from \citet{jarrett2013}:
\begin{equation}
\text{SFR}_{\text{IR}} \left(M_\odot~\text{yr}^{-1}\right) = 7.50 \times 10^{-10} \nu L_{\nu, W4} \left(L_\odot\right)
\end{equation}
with $\nu =1.35\times10^{13}$~Hz \citep{jarrett2011}. SFR$_{\text{IR}}$ accounts for the dust-obscured star-formation that is not readily traced by UV emission, and is one component of the total SFR.

The second component of the total SFR is measured by the UV continuum emission. When available, the Atlas of Nearby Galaxies magnitudes are used (\citet{galexatlas}), otherwise \galex \citep{galex} Near-UV (NUV; central wavelength of 2267~\AA) and Far-UV (FUV; central wavelength of 1516~\AA) magnitudes are downloaded from the MAST archive. NGC~3256 does not have a FUV observation, and we use the NUV relations instead. FUV Magnitudes are corrected for foreground absorption using the relation from \citet{galexatlas}:
\begin{equation}
A_{FUV} = 7.9 ~E \left( B-V \right)
\end{equation}
with $E\left(B-V\right)$ from \citet{schlegel}. Absolute magnitudes are calculated using v4.2 of the IDL package {\it kcorrect} \citep{kcorrect}, assuming H$_0$ = 70~km~s$^{-1}$~Mpc$^{-1}$. Assuming a Kroupa IMF, SFR$_\text{UV}$ is calculated using the relation from \citet{kennicutt2012}:
\begin{equation}
\log{\text{SFR}_{\text{UV}}} = \log{\left(\nu L_\nu\left(L_\odot\right) - 9.7\right)}
\end{equation}
SFR$_{\text{UV}}$ is then combined with SFR$_{\text{IR}}$ to give a total SFR. The combination accounts for dust heating by evolved stars according to the relation from \citet{buat11}:
\begin{equation}
\text{SFR}_{\text{TOT}} = \text{SFR}_{\text{UV}} + 0.83 ~ \text{SFR}_{\text{IR}}
\end{equation}
 This gives a total SFR from both dust-obscured and non-obscured star-formation, with an uncertainty of 20\% \citep{jarrett2013}.

\subsection{Star-formation Rate Surface Densities}
\label{sfrsd}
The star-formation rate surface density (\sfrsdp) has been used as a indicator for galactic outflows because it measures the concentration of the star-formation rate. In areas of concentrated star-formation, supernovae overlap and the energy deposited into the ISM amplifies. If the concentration of the star-formation is high a substantial amount of energy and momentum will be deposited into the gas: the galaxy will drive gas outward in a galactic outflow. This naturally leads to the idea of a threshold of \sfrsd needed to drive galactic outflows \citep{mckee, heckman02, strickland04}. With this picture in mind, it is clear that a local \sfrsd measurement is critical for diagnosing outflows, since star formation within a small area -- not the global \sfrsd -- will drive gas outwards. Unfortunately with the large range of scales probed here, we cannot probe \sfrsd on a local scale for all galaxies. 

We estimate the \sfrsd by the relation:
\begin{equation}
\Sigma_\text{SFR} = \frac{L_\text{COS}}{L_\text{GALEX}} \frac{\text{SFR}_\text{TOT}}{A_\text{COS}}
\end{equation}
Where $\text{SFR}_\text{TOT}$/$A_\text{COS}$ calculates the \sfrsd assuming the entire SFR from the galaxy is within the COS aperture, and L$_\text{COS}$/$L_\text{GALEX}$ (L$_{\text{COS}}$ is measured from the COS spectra with a line free region between 1310 and 1330~\AA) accounts for the fraction of (UV) SFR within the COS aperture. The area probed by the 2.5\rq\rq{} COS aperture (A$_\text{COS}$) is defined by the angular-diameter distance relation.

This estimation of the \sfrsd has a few limitations. First, the projected size of the COS aperture covers a large range of values, depending on the redshift of the galaxy. The areas range from 0.001~kpc$^{-2}$ to 80~kpc$^{-2}$, and samples different physical scales within the galaxy. For low redshift galaxies the \sfrsd does not sample the global \sfrsd of the galaxy, while for higher redshift galaxies \sfrsd does not sample the local variations. For low redshift galaxies the global \sfrsd will be over-estimated because it assumes a high level of star-formation over the entire disk. This method also assumes that the SFR$_{\text{IR}}$ traces the SFR$_{\text{UV}}$, an assumption that works best for unobscured galaxies.

\subsection{Inclinations and Morphologies}
\label{incline}

The inclination is critical to understanding the orientation of the outflowing gas. Since outflows are expected to only cover a fraction of the galaxy, the inclination angle of the galaxy is important for modeling the outflow geometry. \citet{chen10} find a strong positive trend between inclination of the galaxy, the outflow velocity and the outflow equivalent width, while this trend may be less important in this sample because the galaxies are selected to be UV bright (i.e. low obscuration of the young star clusters). Inclinations of the galaxies are calculated from the exponential axis ratios of the r-band SDSS images and the relation in table 8 of \citet{padilla}. When SDSS axis ratios are not available, 2MASS axis ratios are used. While the inclination is measured for compact and irregular galaxies, the interpretation is limited, and we caution that the inclination of these galaxies is not as physically motivated as the spirals. The inclination is only physical for the spirals (14\% of the sample), and has limited applications for the rest of the sample. The errors on the inclinations are estimated to be 20\% \citep{padilla}.

Additionally, morphologies are important to diagnose the merger state of a galaxy. Significant events, like mergers, can temporarily boost the SFR of a galaxy, consequently the merger fraction is important to understand in the context of outflows. Morphologies are assigned with by-eye classification from the SDSS images, literature HST images \citep{overzier09}, or from archived NED classifications. We allow for four different morphologies: Spiral, Merger, Irregular and Compact. A merger classification requires evidence for a merger or interaction: double cores, tidal distortions, or other associated merger remnants. The compact objects are too distant, or too small, to be resolved by the SDSS, and do not have further HST imaging. The sample contains a total of 10 spirals, 16 irregulars, 10 mergers and 15 compact objects.

\subsection{Summary}

The SFRs and stellar masses of the COS sample are calculated using \wise and \galex data. \autoref{fig:masssfr} shows the range of the host galaxy properties, and how they compare to the full SDSS sample \citep{kauffmann2003, brinchmann2004}. At the high mass end, the sample deviates from the SDSS  star-forming \lq\lq{}main-sequence\rq\rq{} by an order of magnitude, and covers four orders of magnitude in both SFR and \mstarp. As a sample, the SFR and \mstar are linearly correlated (SFR~ $\propto$~M$_\ast^{1.01}$). Additionally, we derive \sfrsd from a combination of the COS data and the previously calculated SFRs. Inclinations and morphologies are derived from the SDSS, or from NED.  All values calculated here are tabulated in \autoref{tables}. We have now calculated the relevant properties that we will use to study trends between galactic outflows and their host galaxies.

\section{RESULTS}
\label{results}

Here, we present the relationships between host galaxy properties and the \siii outflows. Since the results heavily rely on how the trends are calculated, we briefly discuss the statistical techniques used to diagnose the trends. For the regression technique, we use the IDL package LTS\_LINEFIT \citep{cappellari}, which is a Robust Least-Squares form of regression that employs a Least Trimmed Squares technique \citep{rousseeuw}. LTS\_LINEFIT finds a subset of the data that globally minimizes the sum of squares among all possible subsets, and is robust down to subsets of half the total sample size \citep{cappellari}. This method is ideal because 1) it accounts for heteroscedastic errors in both the x and y variables; 2) it accounts for, and clips, significant outliers; and 3) it measures the intrinsic scatter within the relationship. We use the Kendall $\tau$ value of the full, unclipped sample, to measure the degree of correlation. Kendall\rq{}s $\tau$ is a non-parametric rank coefficient, with values ranging between -1 (anti-correlation) and +1 (perfect correlation), where a value of 0 signifies no correlation. The Kendall $\tau$ hypothesis test produces a p-value which indicates the significance level of the correlation. The significance level is normalized to the standard deviation ($\sigma$) of a hypothetical test distribution. We consider trends to be highly significant if the significance is greater than 3$\sigma$, and list the highly significant trends as the highest integer value of $\sigma$. For insignificant trends we list the highest half integer $\sigma$ value.  

In \autoref{tab:cor} we tabulate many of the trends of interest. In the first two columns of \autoref{tab:cor} we list the variables, followed by the fit parameters and the measured intrinsic scatter in the third, forth and fifth columns, respectively. The Kendall $\tau$ values, and their corresponding significances, are listed in the last two columns. Since the gap between SBS~1415+437 and the rest of the sample is so large, it may have higher leverage on the fitted relations than the other galaxies, especially if the relations have large intrinsic scatter. Accordingly, we fit relations to the sample with (\autoref{tab:cor}) and without SBS~1415+437 (\autoref{tab:cor_red}). The differences are typically within the errors, except for two cases noted below. We study the trends between host properties and  outflow velocity (\autoref{veltrend}), column density (\autoref{coltrend}), and mass outflow rate (\moutp; \autoref{mouttrend}).

\subsection{Relationship Between Velocity and Galaxy Properties} 
\label{veltrend}

\begin{figure*}
\includegraphics[width = 1.0\textwidth]{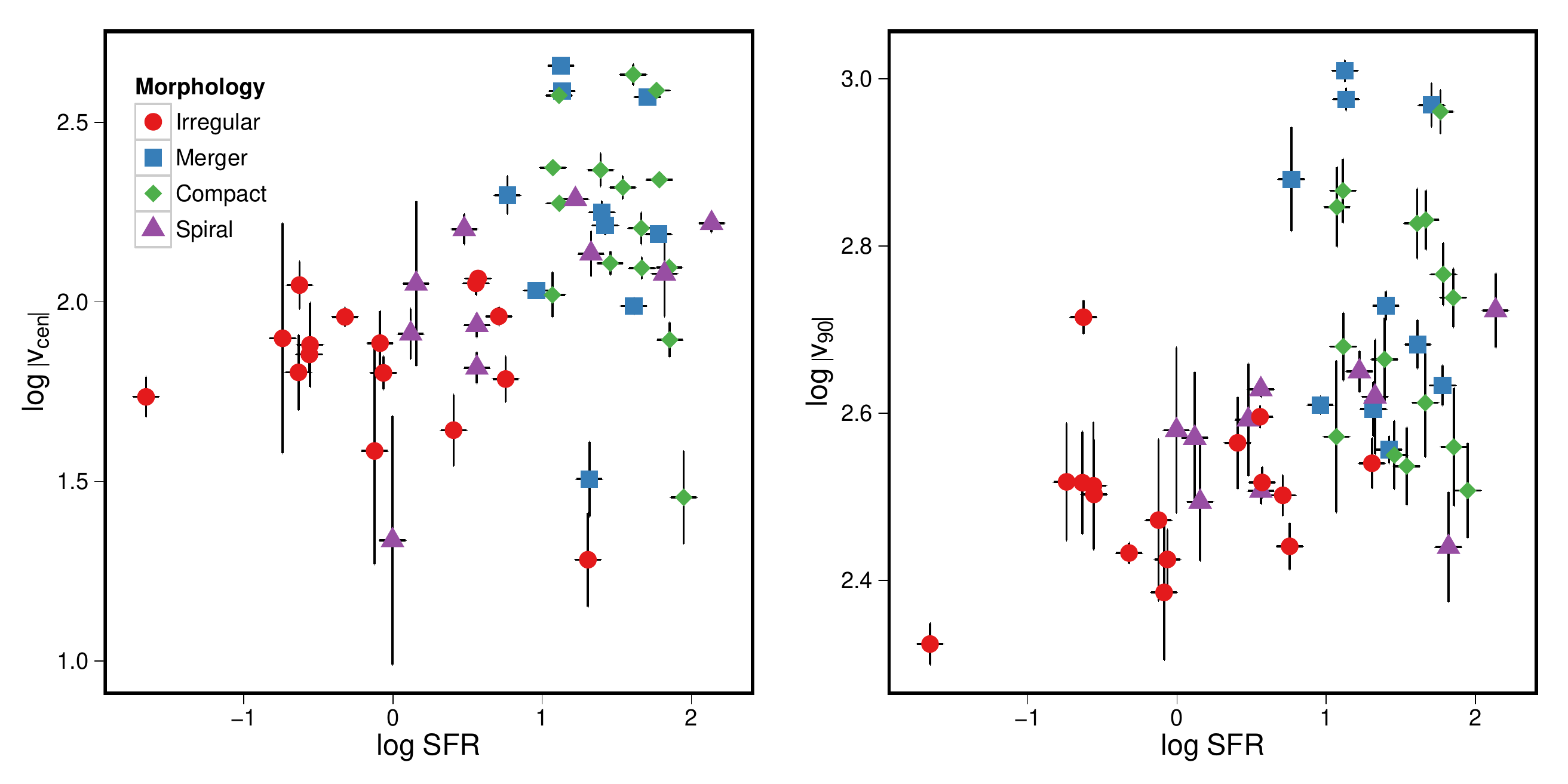}
\caption{Logarithmic relationship between the centroid velocity (\vcenp; on the left, in units of \kmsp), the velocity at 90\% of the continuum (\vnp; on the right, in units of \kmsp), and SFR (in units of M$_\odot$ yr$^{-1}$). Error bars show the 1$\sigma$ error of the velocity and SFR measurements. Point symbols and colors designate the morphology of the galaxy, as labeled by the legend in the left panel. While many galaxies are classified as compact from the SDSS imaging, galaxies above the overall trend may show signatures of mergers or interactions. 3$\sigma$ relationships are seen for both, with scalings given in \autoref{tab:cor}.}
\label{fig:vsfr}
\end{figure*}

The outflow velocity (\voutp) of warm gas may help answer important theoretical questions. What is the energy and momentum deposition of high mass stars? What is the ultimate fate of the gas launched out of galaxies? Will this gas pollute the IGM, or rain back down on the galaxy to promote subsequent star-formation? To answer these questions requires knowledge of the scaling of the outflow velocity and host properties.

First, we study the relationship between \vout and SFR, shown in \autoref{fig:vsfr}. A rise in \vout with SFR is seen up to a SFR of $\sim$10~\sfrp, at which point the scatter in the relationship increases for both \vcen and \vnp. This scatter is mainly driven by points above the main trend with either a merger or compact morphology. This hints that mergers play an important role in generating large velocity outflows (discussed below). \autoref{fig:vsfr} also has three points below the trend for the \vcenp-SFR relationship. These points have complicated profiles, with a strong zero-velocity component, and a higher velocity tail (SDSS J0938+5428 in \autoref{fig:fits} is an example). The zero-velocity absorption decreases \vcenp, but does not impact the measured \vnp, as shown in the left panel of \autoref{fig:vsfr}. This is a limitation of the single line fitting approach. Accounting for the points off the main relation (which are clipped in the regression fitting), a significant trend emerges between \vout and SFR. The Kendall\rq{}s $\tau$ value for \vcen and \vn relations are  0.33 and 0.35 respectively (see \autoref{tab:cor}), and both are correlated at the 3$\sigma$ significance level. \vcen and \vn have significant, albeit shallow, relationships with SFR, with exponents of 0.14 and 0.08 respectively. \autoref{tab:cor_red} illustrates that the fits are consistent within 1$\sigma$ whether SBS~1415+437 is included, or not.

\begin{figure*}
\includegraphics[width = 1\textwidth]{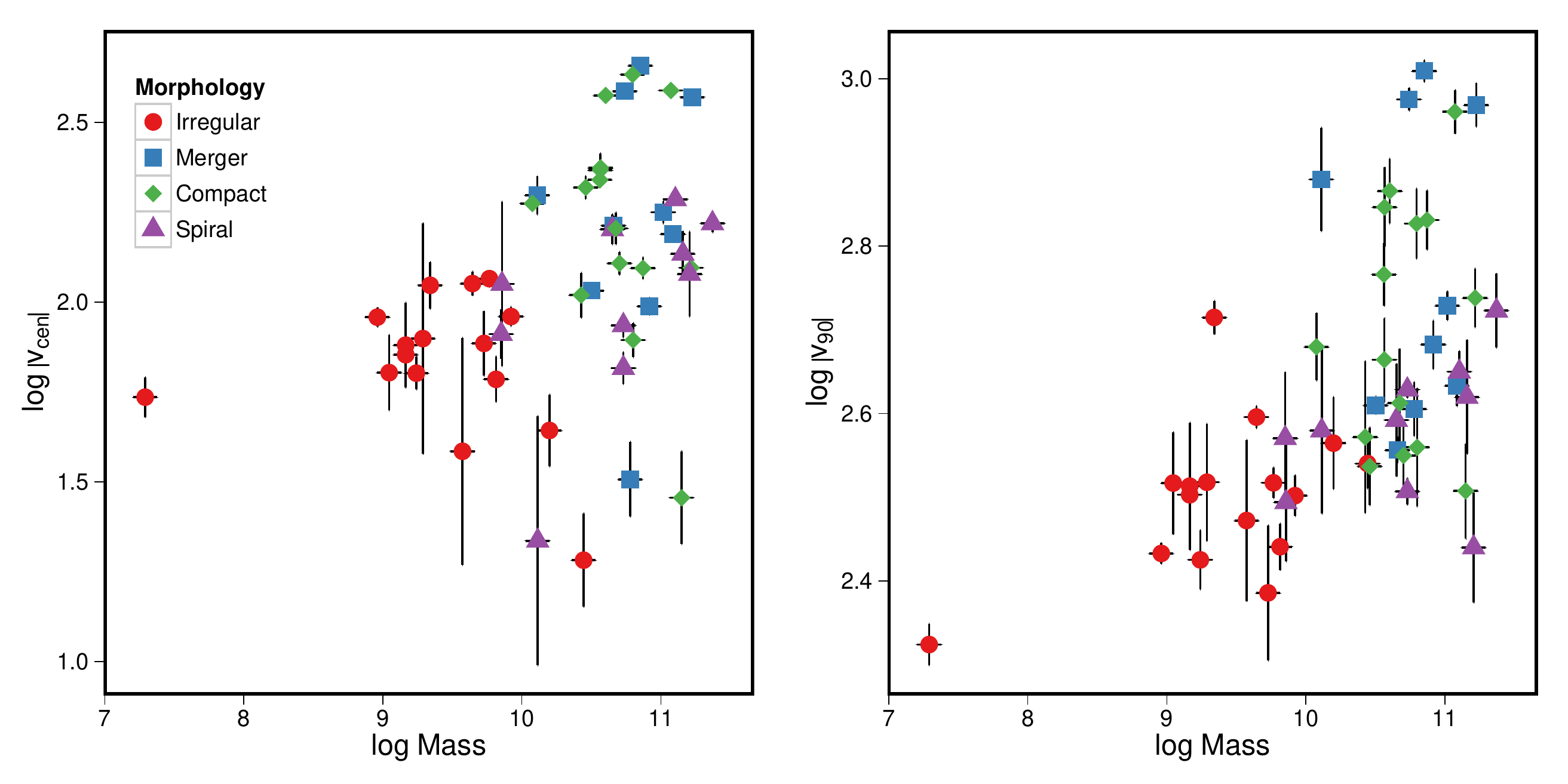}
\caption{Logarithmic relationship between \vcen (left panel, in units of \kmsp), \vn (right panel, in units of \kmsp), and M$_\ast$ (in M$_\odot$). Symbols and colors denote the morphology of the galaxy, as given in the legend. The strong (3$\sigma$ and 4$\sigma$ for \vcen and \vn, repsectively), yet shallow, trend has coefficients given  by \autoref{tab:cor}. If the high velocity, unresolved points are mergers (as argued in the text), a correlation between merger morphology and \vout is seen.}
\label{fig:vmass}
\end{figure*}

The second galaxy property that we study is \mstarp. \autoref{fig:vmass} demonstrates the relationship between \vout and \mstarp. The trends between \vout and \mstar are similar to the trends in SFR, with significance levels of 3 and 4$\sigma$, for \vcen and \vn respectively. When excluding SBS~1415+437, the difference in the trend for \vn and \mstar is larger than 1$\sigma$ (\autoref{tab:cor_red}), and we use the trend without SBS~1415+437 to avoid giving more leverage to SBS~1415+437. While the trends between \vout and \mstar are similar to SFR, the \mstar relations show lower scatter, but higher uncertainty on the estimated coefficients. 

\begin{figure*}
\includegraphics[width = 1.0\textwidth]{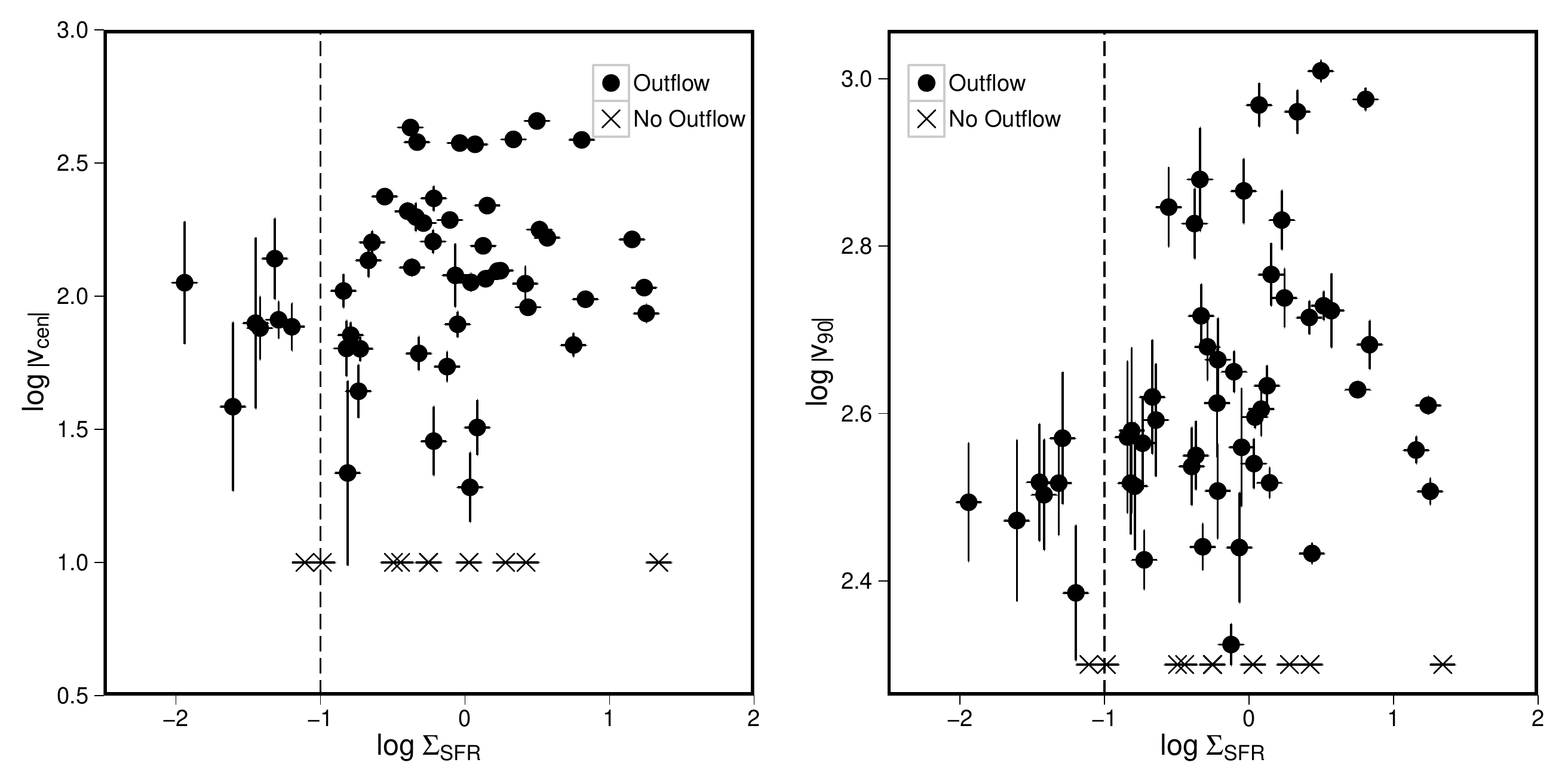}
\caption{Logarithmic relationship between \vcen (in \kmsp), \vn (in \kmsp), and \sfrsd (in M$_\odot$ yr$^{-1}$ kpc$^{-2}$). A weak (2.5$\sigma$) and flat relationship is present. The dot-dashed line denotes \sfrsdp~=~0.1~M$_\odot$~yr$^{-1}$~kpc$^{-2}$, the canonical value above which galactic outflows are driven \citep{heckman02} At these low \sfrsdp, galaxies drive outflows with velocities of 78~\kms. Galaxies not classified as hosting outflows are included in the bottom of the plots as X\rq{}s, with \vcen and \vn artificially set to 10~\kmsp, and 158~\kmsp, respectively.}
\label{fig:vsfrsd}
\end{figure*}

\begin{figure*}
\includegraphics[width = 1.0\textwidth]{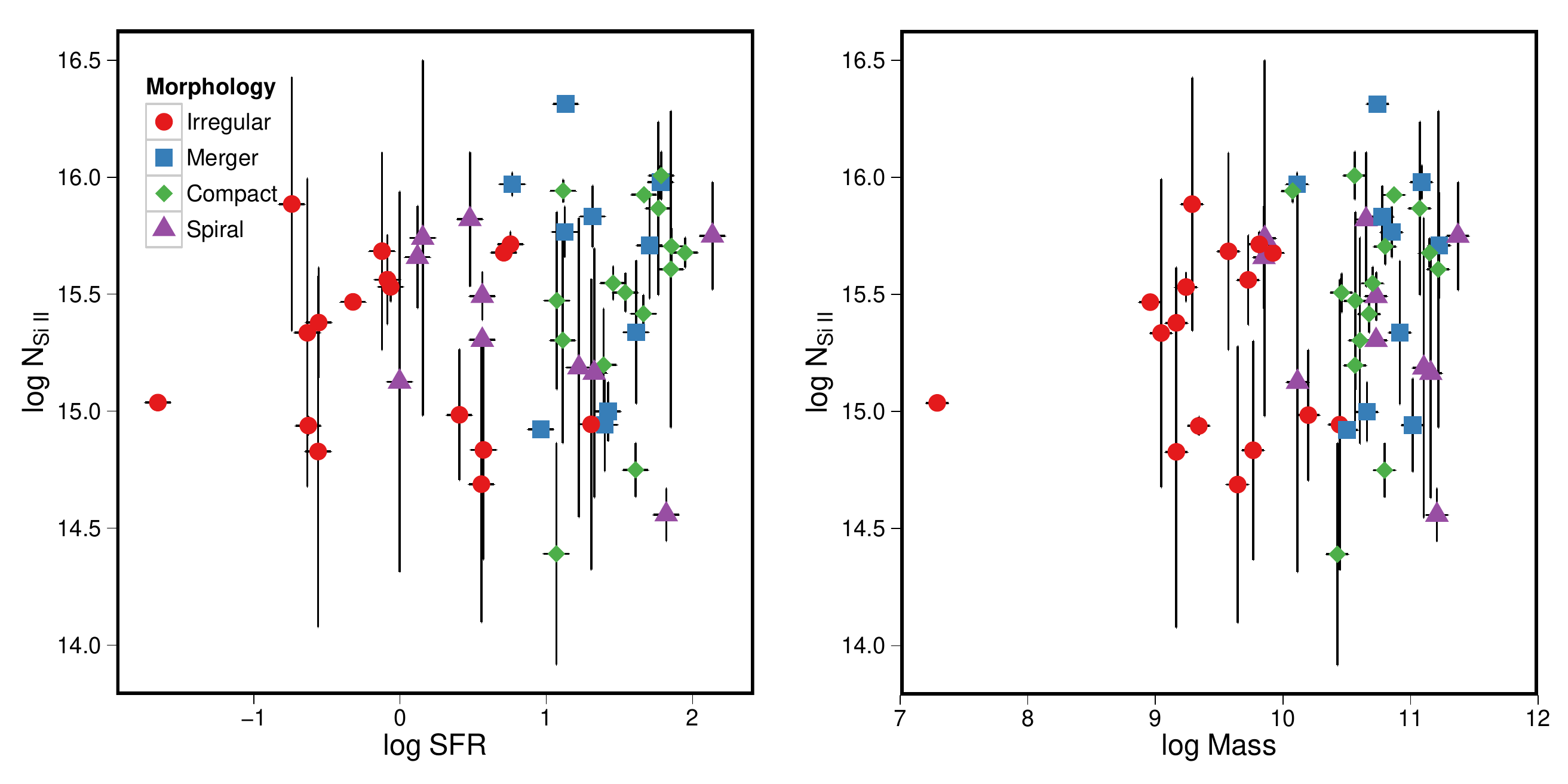}
\caption{Logarithmic relationship between \ns (in cm$^{-2}$), SFR (left panel, in units of M$_\odot$ yr$^{-1}$), and \mstar (right panel, in units of \mstarp). The symbols and colors give the morphology of the galaxy, as shown in the legend. No strong trend is found between these parameters. }
\label{fig:col}
\end{figure*}

The influence of the morphology on \vout is an interesting question; unfortunately, with only 10 spirals and 10 mergers the data cannot produce robust conclusions. We preform a categorical regression and find that mergers generate 0.17~$\pm$~0.07~dex larger \vn than spirals, but only at the 1.75$\sigma$ significance. The low significance is driven mainly by the low number statistics. There are 11 galaxies that lie well above the \vnp-\mstar trend, 6 of which have compact morphologies: five from the Heckman 13017 proposal, and one from the Henry 12928 proposal. The five Heckman galaxies are local Lyman break analogs (LBAs), and previous HST morphological studies show that LBAs are typically undergoing intense mergers or interactions \citep{overzier08, overzier09, overzier11}. Likewise, the Henry proposal targets Green Peas, and, in limited HST imaging, Green Pea galaxies often have strong merger signatures \citep{overzier08, cardamone}. Therefore, it is probable that galaxies off the \vout relation are actually undergoing mergers and interactions. The significance of the merger-\vn relation would be much higher if the 6 compact, high \vout and high \mstar points would be mergers. A higher significance would mean that galaxies undergoing interactions produce faster outflows than galaxies without interactions. However, not all mergers lie off the trend, in fact 60\% of the galaxies with interaction signatures lie on the main trend. In \autoref{escape} we discuss this further.

While \mstar and SFR both exhibit strong correlations with \voutp, we find no significant correlations between \voutp, \sfrsdp, inclination, and sSFR (SFR/\mstarp). \autoref{fig:vsfrsd} shows the weak (2.5$\sigma$), flat (\vn~$\sim \Sigma_{\text{SFR}}^{0.05}$) relationship between \vout and \sfrsdp. Additionally, the null correlation between inclination and \vout most likely arises from the dearth of spiral galaxies in the sample, because inclination is most physically meaningful for spirals. These null correlations imply that SFR and \mstar account for most the variation in \vout from the measured parameters. Due to the strong linear correlation between SFR and \mstar the distinction between the two trends cannot be definitively made. A multivariate fit does not produce a more statistically significant relation than the individual fits.  Meanwhile, a principle component analysis (PCA) finds the angle between the loadings of \vn (and \vcenp) and \mstar to be slightly less than between \vn and SFR. However, the difference is less than 1$^{\circ}$, and does not significantly distinguish the principle drivers of the trend.

\begin{deluxetable*}{cc|ccccc}
\tablewidth{0pt}
\tablecaption{Measured Relations}
\tablehead{
\colhead{y} &
\colhead{x} &
\colhead{slope} &
\colhead{Intercept} &
\colhead{Scatter (dex)} &
\colhead{Kendall\rq{}s $\tau$} &
\colhead{Significance}
}
\startdata
log(\vcenp) & log(SFR)  &  0.14 $\pm$ 0.03  & 1.94 $\pm$ 0.03  & 0.14 $\pm$ 0.02 & 0.33 & 3$\sigma$ \\
log(\vnp) & log(SFR)  &  0.08 $\pm$ 0.02 & 2.50 $\pm$ 0.08  &0.06 $\pm$ 0.01  & 0.35  & 3$\sigma$ \\
log(\vcenp)  & log(\mstarp)   & 0.12 $\pm$ 0.02 &  0.76 $\pm$ 0.22   & 0.09 $\pm$ 0.04 & 0.32  & 3$\sigma$ \\
log(\vnp)  & log(\mstarp)   & 0.08 $\pm$ 0.01 &  1.7$\pm$ 0.06   & 0.00 $\pm$ 0.00 & 0.42  & 4$\sigma$ \\
log(\vnp)  & log(\sfrsdp)   & 0.05 $\pm$ 0.02 & 2.58 $\pm$ 0.01   & 0.08 $\pm$ 0.01  & 0.28  & 2.5$\sigma$ \\
log(\nsp)  & log(\mstarp)   & 0.08 $\pm$ 0.02 &  14.71 $\pm$ 0.2  &  0.00 $\pm$ 0.00& 0.13 & 1$\sigma$ \\
log(\nsp)  & log(SFR)   & 0.22 $\pm$ 0.02 &  15.5 $\pm$ 0.02  & 0.00 $\pm$ 0.00 & 0.15  & 1.5$\sigma$ \\
log(\moutp)  & log(SFR)   & 0.46 $\pm$ 0.09 &  0.17 $\pm$ 0.11  &  0.48 $\pm$ 0.08& 0.45 & 4$\sigma$ \\
log(\moutp)  & log(\mstarp)   & 0.42 $\pm$ 0.10 &  -3.8$\pm$ 1.10 & 0.52 $\pm$ 0.08 & 0.39 & 4$\sigma$ \\
log(\moutp)  & log(\sfrsdp)   & 0.25 $\pm$ 0.13 & 0.59 $\pm$ 0.09  & 0.57 $\pm$ 0.01  & 0.21& 2$\sigma$ \\
log($\eta$)  & log(\mstarp)   & -0.58 $\pm$ 0.1 &  5.65 $\pm$ 0.98  & 0.47 $\pm$ 0.02 &-0.35 & 3$\sigma$ \\
log(\vn)  & log(\vcirc)   & 0.37 $\pm$ 0.03 &  1.7 $\pm$ 0.06  & 0.0 $\pm$ 0.00 & 0.41 & 4$\sigma$ \\
log(\vcenp)  & log(\vcircp)   & 0.56 $\pm$ 0.1 &  0.83 $\pm$ 0.21  & 0.09 $\pm$ 0.04 & 0.32 & 3$\sigma$ \\
log($\eta$)  & log(\vcircp)   & -2.60 $\pm$ 0.43 &  5.35 $\pm$ 0.93  & 0.47 $\pm$ 0.02 & -0.35 & 3$\sigma$ \\
\enddata
\tablecomments{\ Least Trimmed Squares Robust Regression coefficients of the various trends studied. The two left columns give the variables studied, while the third and fourth columns give the slope (exponent) and intercept (constant) of the fit. The scatter (fifth column) is the intrinsic scatter of the relation, after accounting for errors in both x and y. If all of the scatter can be attributed to measurement errors (as in the \ns relations) the scatter is reported as 0.0 $\pm$ 0.00. Kendall\rq{}s  $\tau$ is a measure of correlation between the two variables, with +1 implying a perfect correlation, and zero implying no correlation. The significance level is for the Kendall\rq{}s $\tau$ test. We consider trends to be highly significant if the significance is greater than 3$\sigma$.}
\label{tab:cor}
\end{deluxetable*}

\begin{deluxetable*}{cc|ccccc}
\tablewidth{0pt}
\tablecaption{Measured Relations without SBS~1415+437}
\tablehead{
\colhead{y} &
\colhead{x} &
\colhead{slope} &
\colhead{Intercept} &
\colhead{Scatter (dex)} &
\colhead{Kendall\rq{}s $\tau$} &
\colhead{Significance}
}
\startdata
log(\vcenp) & log(SFR)  &  0.20 $\pm$ 0.04  & 1.95 $\pm$ 0.05  & 0.20 $\pm$ 0.02 & 0.33 & 3$\sigma$ \\
log(\vnp) & log(SFR)  &  0.08 $\pm$ 0.02 & 2.51 $\pm$ 0.08  &0.06 $\pm$ 0.01  & 0.35  & 3$\sigma$ \\
log(\vcenp)  & log(\mstarp)   & 0.14 $\pm$ 0.03 &  0.63 $\pm$ 0.28   & 0.10 $\pm$ 0.04 & 0.32  & 3$\sigma$ \\
log(\vnp)  & log(\mstarp)   & 0.11 $\pm$ 0.01 &  1.4$\pm$ 0.11   & 0.02$\pm$ 0.01 & 0.42  & 4$\sigma$ \\
log(\vnp)  & log(\sfrsdp)   & 0.05 $\pm$ 0.02 & 2.58 $\pm$ 0.01   & 0.08 $\pm$ 0.01  & 0.28  & 2.5$\sigma$ \\
log(\nsp)  & log(\mstarp)   & 0.08 $\pm$ 0.02 &  14.71 $\pm$ 0.51  &  0.00 $\pm$ 0.00 & 0.12 & 1$\sigma$ \\
log(\nsp)  & log(SFR)   & 0.22 $\pm$ 0.02 &  15.53 $\pm$ 0.02  & 0.00 $\pm$ 0.00 & 0.14  & 1$\sigma$ \\
log(\moutp)  & log(SFR)   & 0.58 $\pm$ 0.09&  0.11 $\pm$ 0.11  &  0.43 $\pm$ 0.02 & 0.45 & 4$\sigma$ \\
log(\moutp)  & log(\mstarp)   & 0.56 $\pm$ 0.12 &  -5.2$\pm$ 1.30 & 0.51 $\pm$ 0.07 & 0.39 & 4$\sigma$ \\
log(\moutp)  & log(\sfrsdp)   & 0.25 $\pm$ 0.13 & 0.60 $\pm$ 0.09  & 0.57 $\pm$ 0.01  & 0.22 & 2$\sigma$ \\
log($\eta$)  & log(\mstarp)   & -0.53 $\pm$ 0.12 &  5.10 $\pm$ 1.3  & 0.49$\pm$ 0.01 &-0.32 & 3$\sigma$ \\
log(\vnp)  & log(\vcircp)   & 0.49 $\pm$ 0.05 &  1.5 $\pm$ 0.1  & 0.02 $\pm$ 0.01 & 0.39 & 4$\sigma$ \\
log(\vcenp)  & log(\vcircp)   & 0.61$\pm$ 0.12 &  0.7 $\pm$ 0.3 & 0.10 $\pm$ 0.04 & 0.30 & 3$\sigma$ \\
log($\eta$)  & log(\vcircp)   & -2.36 $\pm$ 0.54 &  4.80 $\pm$ 1.2  & 0.49 $\pm$ 0.01 & -0.32 & 3$\sigma$ \\
\enddata
\tablecomments{Same as \autoref{tab:cor}, but when SBS~1415+437, the lowest mass galaxy in the sample, is excluded from the sample. We calculate both situations because SBS~1415+437 has a large separation from the rest of the sample, and may have higher leverage than other individual galaxies. The only relationships that change by more than 1$\sigma$ with the inclusion of SBS~1415+437 are the  \vn~$\propto$~\mstar and \vn~$\propto$~\vcirc relationships. For the \vn~$\propto$~\mstar and \vn~$\propto$~\vcirc trends we use the relations tabulated here, while for the other relations we use the results from \autoref{tab:cor}.}
\label{tab:cor_red}
\end{deluxetable*}

\subsection{Relationship Between Column Density and Galaxy Property}
\label{coltrend}

We use column density (\nsp) as a proxy for the amount of \siii in the outflow. When assumptions about the geometry and metallicity are combined with \nsp, the average mass outflow rate (\moutp) can be calculated (see \autoref{mouttrend}). Column densities will also be used in subsequent work to model the ionization structure of the galactic outflow.

\autoref{fig:col} shows how \ns relates to SFR, and \mstarp. The trends are shallow (exponents of 0.22 and 0.24 for SFR and \mstarp, see \autoref{tab:cor}) and insignificant (\autoref{tab:cor}). Large uncertainties on some of the column densities make possible trends difficult to discern. Unlike the \vout trends, we do not find significant trends between \nsp, \mstarp, and SFR.

\subsection{Relationship Between \mout and Galaxy Properties} 
\label{mouttrend}

The mass outflow rate (\moutp) is the mass per unit time that escapes the star-forming region. \mout regulates the pace of star-formation by removing gas from the star-forming region that otherwise could gravitationally collapse to form stars. We do not directly measure \mout from the COS data, rather we must model the outflow geometry to derive the \moutp. The outflow geometry is uncertain: theoretical simulations predict that warm gas can be found in either thin shells \citep{cooper, creasey}, or in clumpy filaments \citep{hopkins12b}. Unfortunately, observations do not resolve the issue either. \citet{rubin11} detect Fe~{\sc II$^\ast$} emission out to at least 7~kpc from the starburst in TKRS4389; indicative of an extended outflow structure. Meanwhile, the nearby starburst M82 has clumpy H$\alpha$ knots and filaments within 1~kpc from the starburst \citep{shopbell, westmoquette}. The geometry and morphologies of outflows are very uncertain, and an important outstanding question of galaxy evolution.

We follow \citet{weiner} and assume the observed \siii resides in a thin shell of radius R.  Typically, studies fix the radius of the shell -- normally, at 5~kpc \citep{rupke2005b, weiner} -- to measure \moutp. However, if we assume the outflows are powered by the energy released from star-formation, more energy deposited into the ISM would launch the gas to a larger radius. This means that R depends on the SFR of the galaxy, which is important to include when the sample covers 4 orders of magnitude in SFR. \citet{grimes2005}  find the radius enclosing 90\% of the X-ray emission (assumed to be the outflow radius) to depend on the SFR and \mstarp. Using their radius and L$_{IR}$ data, we derive a 4$\sigma$ R-SFR relationship, with the far-IR SFR calibrations \citep{kennicutt2012}, to be:
\begin{equation}
R\left(\text{kpc}\right) = \left(4.4 \pm 0.7\right) \left(\frac{\text{SFR}}{10~M_\odot~\text{yr}^{-1}}\right)^{0.34  \pm .05}
\end{equation}
The relation provides a radius of 8.9~kpc for the galaxy TKRS4389 (mentioned above), consistent with the lower limit of 7~kpc \citep{rubin11}. This relations produces a median outflow radius of 4.7~kpc for the sample, with values ranging between 0.6 and 10.6~kpc.

Along with the outflow radius, the outflow covering fraction is an uncertain geometric parameter. The covering fraction depends on the clumping of the outflow, the opening angle of the outflow, the exact morphology of the outflow (biconical, or spherical), the radial extent of the outflow, and the relative size of the star-forming region to the outflow, to name a few. Following \citet{rupke2005b}, we divide the covering fraction into a product of two quantities: one describing the geometry of the wind ($C_G$), and one describing the clumping of the wind ($C_f$). From the detection rate of $\sim$80\% \citet{rupkee2005} estimate $C_G$ to be roughly 0.8. Considering our similar detection rate, we use a constant $C_G = 0.8$. This quantity may vary with SFR or \mstar \citep{rupke2005b}, but we do not have sufficient data to produce a trend. For the clumping of the wind ($C_f$) we use the covering fraction measured from the absorption line fit (see \autoref{data}). These covering fractions vary between 0.34 and 1.0, with a median of 0.9. We do not observe a significant trend between C$_f$ and host galaxy properties. Once the geometry of the wind is modeled, we can include the other wind parameters.

The total mass of warm Hydrogen, in solar units, is calculated by assuming that the majority of the gas is warm and partially ionized. We convert \ns into a hydrogen column by dividing by the best-fit metallicity, derived in \autoref{data}, and the local relative abundance, including dust-depletion, of \siii to H in the Warm Ionized Medium (WIM): $\left[\text{Si/H}\right]_{\text{gas}} = 3.1 \times 10^{-5}$ \citep{jenkins, draine}. We caution that although we use the best-fit metallicity for the galaxy, the actual metallicity of the outflowing gas is uncertain. Complete modeling of the metallicity, with all the UV transitions, will be left for future work. Functionally, N$_\text{H}$ is calculated by: 
\begin{equation}
\text{N}_\text{H} = \frac{N_{\text{Si~{\sc II}}}}{\left[\text{Si/H}\right]_{\text{gas}} Z}
\end{equation}
The N$_\text{H}$ values range from 6.6~$\times 10^{18}$ to 5.3~$\times 10^{20}$~cm$^{-2}$, with a median of 5.6~$\times 10^{19}$~cm$^{-2}$, in agreement with N$_\text{H}$ from the DEEP2 survey \citep{weiner} for slightly higher redshift galaxies. We assume no ionization correction. The spectra do have \oi and \siiii present, indicative of both higher and lower excitation gas, but \siii is the dominant ionization species in the WIM, and we make no ionization correction. In future work we will use other transitions to calculate the ionization corrections, and produce total mass outflow rates. Here, these values roughly hold for gas with a temperature between  6$\times 10^3$ and $4\times10^4$~K \citep{mazzotta}.

Combining the geometry, N$_H$, and the velocity of the \siii gas, we calculate the mass outflow rate (\moutp; in units of M$_\odot$~yr$^{-1}$). We use the equation from \citet{weiner}:
\begin{equation}
\text{\mout} \simeq 22 ~\text{M}_{\odot}~\text{yr}^{-1}~ C_G~ C_f~ \frac{N_ H}{10^{20}~\text{cm}^{-2}} ~ \frac{R}{5~\text{kpc}} ~\frac{\text{v}_\text{cen}}{300~\text{km s}^{-1}} 
\end{equation}
If we assume a thick wind instead of a thin shell, the \mout values would change by a factor of 1/3 \citep{weiner}, and only the normalizations would change, not the relations. We use \vcen as the outflow velocity because it measures the bulk velocity of the outflow, and describes the bulk mass outflow rate. We do not have a measure of the column density for the highest velocity gas, therefore we do not calculate a maximum velocity \moutp. The errors for \mout are propagated through according to the measured errors on each parameter. \mout ranges from 0.18~\sfr to 56~\sfrp,  with a median of 3.7~\sfr (compared to a median SFR of 13~\sfrp).

\begin{figure*}
\includegraphics[width = 1.0\textwidth]{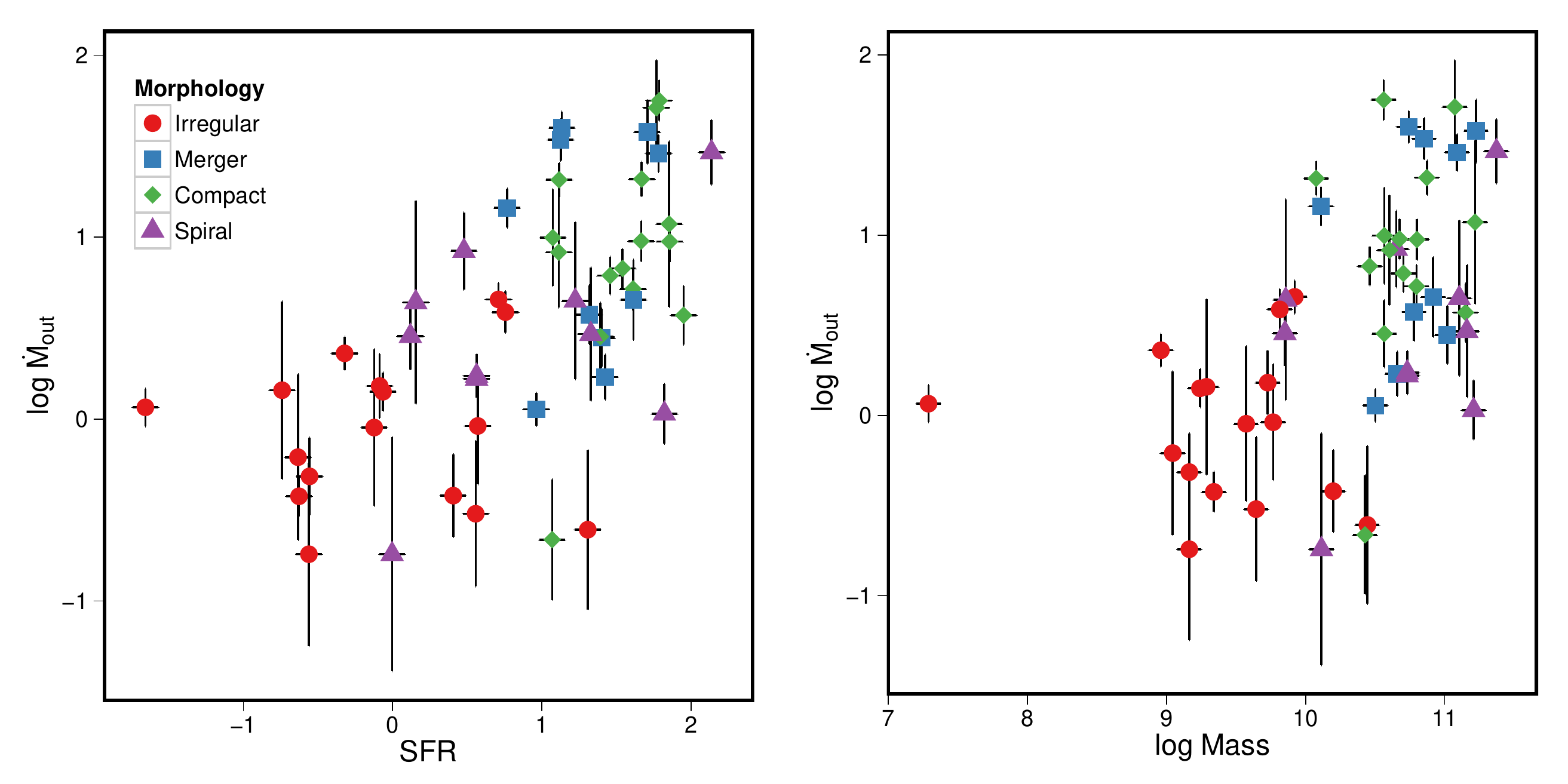}
\caption{Logarithmic relationship between the mass outflow rate (\moutp; in units of \sfrp), SFR (left panel, in units of \sfrp), and \mstar (right panel, units of \mstarp).  The \moutp-\mstar and \mout-SFR correlations are strong (4$\sigma$ and 5$\sigma$ respectively), both with slopes of roughly 1/2.}
\label{fig:mout}
\end{figure*} 

Once \mout is calculated, we can explore the trends of \mout with galaxy properties. \autoref{fig:mout} shows that  \mstar and SFR both exhibit strong correlations with \moutp: \mstar has a Kendall\rq{}s $\tau$ value of 0.46, while SFR has a value of 0.49 (\autoref{tab:cor}). We find power-law relationships between these two quantities with exponents of 0.50 and 0.48 for SFR and \mstarp, respectively. SBS~1415+437 does not affect the fits by more than the 1$\sigma$ errors (\autoref{tab:cor_red}). While the \sfrsd trend is insignificant (2$\sigma$), it has an exponent of 0.25.

The scalings between \mout and host properties show significant scatter (0.5~dex, \autoref{tab:cor}). Small, shallow, secondary correlations between the variables may exist that drive the scatter in these relationships. We explore this issue by preforming a multivariate M-estimator Robust regression fit to the \moutp-SFR-\mstar plane \citep{stats}. We find the multivariate relationship between \moutp, SFR, and \mstar to be
\begin{equation}
\mathsmaller{\dot{M}_o = 6.7~\text{M}_\odot~\text{yr}^{-1}  \left(\frac{\text{SFR}}{10M_\odot \text{yr}^{-1}}\right)^{0.72 \pm .18} \left(\frac{M_\ast}{10^{10}M_\odot}\right)^{-0.33 \pm 0.19}}
\label{eq:multimout}
\end{equation}
The overall trend is strong ($4\sigma$), but most of the significance is in the SFR (3$\sigma$). We also explore the possibility of a \sfrsd dependence, but find the multivariate trend with \sfrsd to be statistically flat.

\begin{figure*}
\includegraphics[width = 1.0\textwidth]{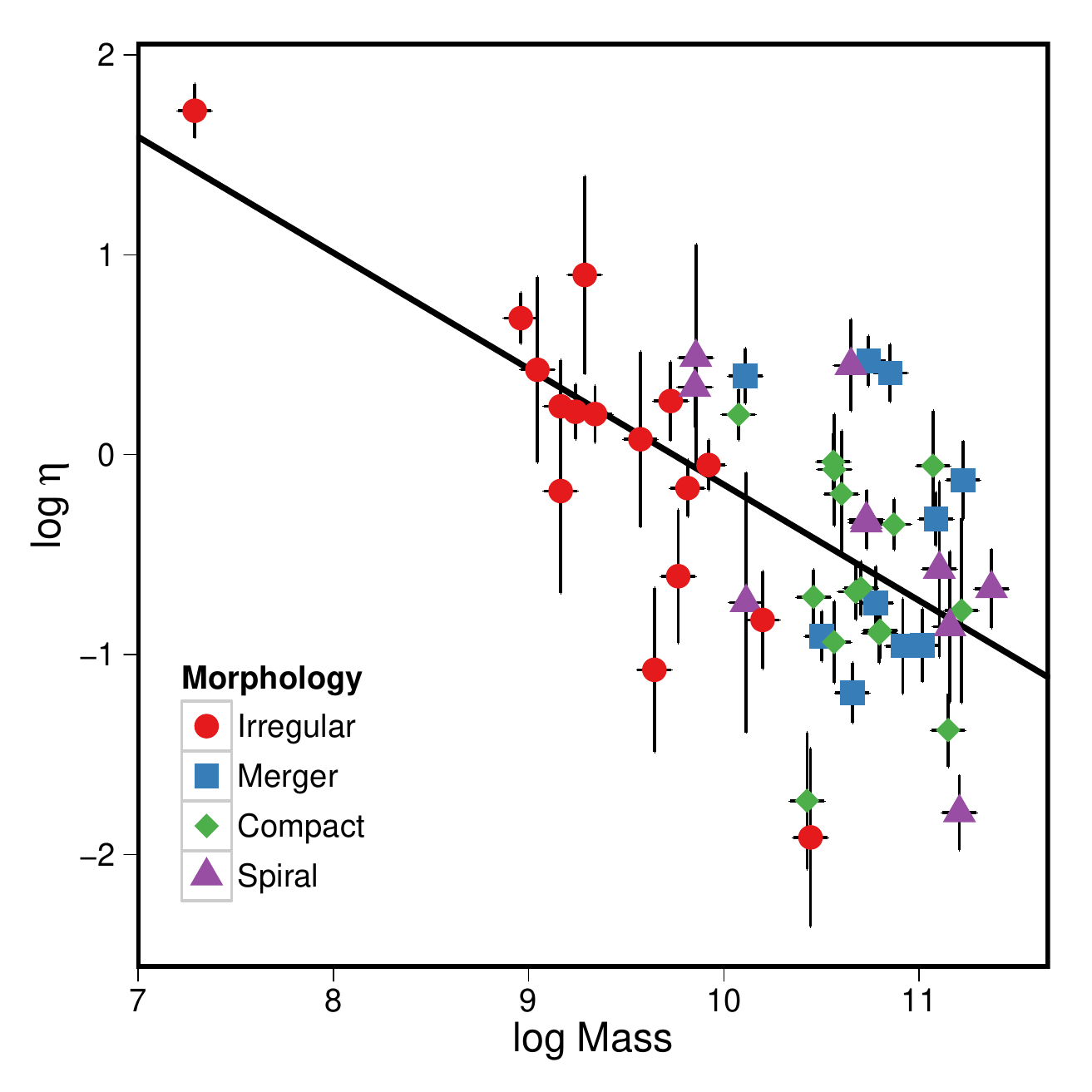}
\caption{Logarithmic relationship between the mass-loading factor ($\eta = $\mout/SFR) and \mstar (in units of M$_\odot$). The regression line is shown as the solid black line with the scaling of $\eta~\propto~$\mstarp$^{-0.58}$. The relation has a Kendall $\tau$ correlation factor of -0.35 (3$\sigma$). }
\label{fig:eff} 
\end{figure*}

The easiest way to parameterize \mout is to normalize \mout by the SFR, and calculate the mass-loading factor ($\eta$).   The mass-loading factor is defined in the literature as: 
\begin{equation}
\eta = \frac{\dot{M_o}}{\text{SFR}}
\label{eq:eta}
\end{equation}
$\eta$ measures how efficiently the galaxy forms stars: a galaxy with a high $\eta$ ejects more mass than it forms stars. Thus, high mass-loading factors arise in galaxies with inefficient star formation. The median $\eta$ for the sample is 0.47, with values between 0.01 and 52.6. These values are roughly consistent with values found for other samples \citep{rupke2005b, rubin13, arribas2014}.  SBS1415+437 is a remarkable case-study: with \mout of  1.2~\sfrp, this galaxy has $\eta = 52.6$. The mass outflow rate is 53 times larger than the SFR. Without an external gas supply, this low mass galaxy (log(\mstarp) = 7.3) will rapidly exhaust the available fuel for star-formation. Looking at the trend of $\eta$ and host galaxy properties, we find that $\eta$ has a strong correlation (3$\sigma$; Kendall\rq{}s $\tau$ = -0.29) with \mstarp, as shown in \autoref{fig:eff}. The relationship takes the functional form of
\begin{equation}
 \eta \simeq 0.74 ~ {\left(\frac{\text{M}_\ast}{1 \times 10^{10} M_\odot}\right)^{-0.58 \pm 0.1}}
\end{equation}
with an appreciable amount of scatter (0.5 dex). There are no significant trends between \sfrsd and $\eta$ ($\tau$ = -0.10, or 1$\sigma$).

Here, we have found significant relations between the \mout and host galaxy parameters. Similar to the \voutp, both SFR and \mstar are of similar scaling, with \mout proportional to SFR$^{1/2}$. There is significant physical scatter in these relations. We find a strong multivariate correlation between \moutp, and SFR$^{3/4}$ M$_\ast^{-1/3}$. We explore the relationship between the mass-loading efficiency factor ($\eta$) and find a strong correlation proportional to M$_\ast^{-1/2}$. These relationships are changed by less than 1$\sigma$ with the inclusion of the lowest mass galaxy, SBS~1415+437.

\section{DISCUSSION}
\label{discussion}

The scaling relations found in \autoref{results} describe the injection of energy into the ISM through star formation. The implications for galactic evolution are fundamental, and we discuss only a few here. First, we present the scaling relations and compare them to previous studies. We then discuss how these relations relate to the outflow driving mechanisms. The energetics are then used to discuss the final fate of the gas, and how this qualitatively, and quantitatively, produces the observed mass-metallicity relation.

\subsection{Scaling Relations}

The scaling relations of \autoref{results} are our primary result. These relations link the characteristics of the star-formation driven outflows to their host galaxies. We find statistically significant scaling relations between \vout (both \vcen and \vnp) and \mout (or $\eta$), with the \mstar and SFR of the host galaxies. Since \mstar and SFR are highly covariant, we cannot make a distinction between these properties, however, in \autoref{energy}  we argue that the energetics favor SFR as the fundamental quantity for the \vout relations. These scaling relations encapsulate the complicated physics of launching galactic outflows, and are given, in terms of SFR and \mstarp, as:
\begin{equation}
\begin{aligned}
\text{v}_\text{cen} &= \left(121 \pm 9~\text{km s$^{-1}$}  \right)  \left( \frac{\text{SFR}}{10~M_\odot~\text{yr}^{-1}}\right)^{0.14 \pm 0.03} \\
\eta &= \left(0.74 \pm 1.7\right)  \left( \frac{M_\ast}{1 \times 10^{10} M_\odot}\right)^{-0.58 \pm 0.1}  \\
\end{aligned}
\label{eq:scaling}
\end{equation}
In terms of \mstar the relation for \vcen is:
\begin{equation}
\text{v}_\text{cen} =\left( 100 \pm 51 ~\text{km s$^{-1}$} \right)  \left( \frac{M_\ast}{1 \times 10^{10} M_\odot} \right)^{0.12 \pm 0.02}
\label{eq:scaling1}
\end{equation}
We use the observed dust corrected, stellar mass Tully-Fisher relation from \citet{bell01} to calculate \vcirc from:
\begin{equation}
\log{\text{M}_\ast} = 0.52 + 4.49 \log{\text{v}_{\text{circ}}}
\label{eq:obstf}
\end{equation}
 \autoref{eq:scaling} can then be recast in terms of \vcirc as: 
\begin{equation}
\begin{aligned}
\text{v}_\text{cen} &= \left(129 \pm 62 ~\text{km s}^{-1}\right) \left(\frac{\text{v}_\text{circ}}{200~\text{km s}^{-1}}\right)^{0.56 \pm 0.1} \\
\eta &= \left(0.23 \pm 0.5\right)  \left(\frac{\text{v}_\text{circ}}{200~\text{km~s}^{-1}}\right)^{-2.6 \pm 0.43} \\
\label{eq:scalingcirc}
\end{aligned}
\end{equation}
Previous studies have found similar relations between \vout and SFR. \citet{rupke2005b} use Na~D absorption to find the scaling of v$_\text{max} \propto \text{v}_\text{circ}^{0.85 \pm 0.15}$ and v$_\text{max} \propto$ SFR$^{0.21 \pm 0.04}$, about a 1.2 and 1$\sigma$ deviation, respectively from \autoref{eq:scalingcirc} and \autoref{eq:scaling}. Likewise, \citet{arribas2014} use H$\alpha$ emission to find the \voutp-SFR scaling of \vout~$\propto$ ~SFR$^{0.24 \pm 0.05}$, a 1.3$\sigma$ deviation from \autoref{eq:scaling}. Meanwhile, \citet{weiner} use Mg~{\sc II} absorption to find the velocity at 25\% of the continuum (v$_{25\%}$) to scale as v$_{25\%} \propto M_\ast^{0.11}$ and SFR$^{0.17}$, within 1$\sigma$ of our results for \vcenp. Slight discrepancies are expected because of the different parameter spaces of the samples and outflow tracers used. \citet{rupke2005b} use the cool gas tracer Na~D to probe the outflows of mainly ultra-luminous infrared galaxies. Since \siii traces warmer gas, that is not as easily destroyed by high energy photons and shocks, it is a more robust tracer of the higher temperature, higher mass flux, warm ionized outflow \citep{hopkins12a}. Meanwhile, \citet{arribas2014} uses H$\alpha$ emission to characterize the outflow properties. Wood et al. (2015, in preparation) show that the broad H$\alpha$ component may probe the shocked gas at the base of the outflow more effectively than the bulk of the outflow, making H$\alpha$ a less reliable tracer of galactic winds. Since \citet{weiner} probe the same temperature regime, the consistency between the relations shows that the scaling depends on the temperature regime probed (as will be discussed in a future work). Our results can also be compared  to previous observed $\eta$ scaling relations. \citet{arribas2014} find the broad H$\alpha$ component to scale as $\eta \propto \text{M}_\ast^{-0.43}$, while \citet{rupke2005b} find $\eta$ to scale with the K-band magnitude as $\eta \propto M_K^{-0.42 \pm 0.17}$, both slightly shallower relations. The results presented here are generally marginally consistent with previous findings from studies.

A large parameter space and sample size, like the one presented here, is essential for calculating these trends. \autoref{fig:vsfr} and \autoref{fig:eff} show that these relations have a large physical scatter (0.14 and 0.47~dex, for the \vcenp-SFR and $\eta$-\mstar relations respectively), with numerous outliers. The scatter may arise from second order correlations, like \autoref{eq:multimout} where the \mout scales as SFR$^{0.72}$ and M$_\ast^{-0.33}$, which adds a second order dependence at fixed SFR. Additionally, outliers can influence the measured trends. In \autoref{escape} we hypothesize that these outliers likely arise from an increase in \vout during the merger sequence. Studies that do not sample the full dynamic range, or do not have a large enough sample size, will be more sensitive to outlier and may report biased scaling relations.  

We can also compare our relations with those found in various galactic hydrodynamic simulations. By including the complicated physics of radiation pressure and supernovae explosions, \citet{hopkins12a} find simulated galactic outflows to scale as $\eta$ $\propto$M$^{-(0.25\rightarrow0.5)}_\ast$ (or v$_\text{circ}^{-\left(1.1 \rightarrow 2.25\right)}$ using the observed Tully-Fisher relation in \autoref{eq:obstf}), a range marginally consistent with our observations. Similarly, \citet{creasey} simulate the effect of supernovae feedback, and find the scaling relation that matches the Tully-Fisher relation to be $\eta$ $\propto$v$_\text{circ}^{-2.5}$. Although still in the nascent phase, simulations of individual galaxies marginally produce similar scaling relations as found here.

The null correlations are equally as important. The role of \sfrsd has been cited as a criterion for launching galactic outflows, such that to launch an outflow a galaxy must have \sfrsd >~0.1~\sfrp~kpc$^{-2}$ \citep{heckman02}. However, we do not find a statistically significant correlation between \sfrsd and any outflow property. Moreover, \autoref{fig:vsfrsd} shows seven galaxies with \sfrsd less than  0.1~\sfrp~kpc$^{-2}$ that drive outflows with a median \vcen of -78~\kmsp, a result consistent with \citet{rubin13}. \sfrsdp, as measured here, does not act as a threshold to drive outflows. Furthermore, in \autoref{fig:vsfrsd} we include galaxies that we characterize as not having an outflow. These galaxies cover a similar parameter space, yet do not produce outflows. This is most likely due to the location of the COS aperture, the observation inclination, the outflow geometry, or the evolutionary state of the outflow. We do not observe a \sfrsd threshold to drive galactic outflows, rather, outflows appear to be prevalent in >~80\% of actively star-forming galaxies down to the limit of our sample, \sfrsd$ = 0.01~M_\odot~\text{yr}^{-1}~\text{kpc}^{-2}$.

We have presented the scaling relations between host galaxy properties and warm gas outflows. With these relations, we now explore the energetics of the outflows.

\subsection{Outflow Energetics}
\label{energy}
The outflow energetics link star formation feedback and outflow properties. High mass stars release energy into the ISM through supernovae explosions, stellar winds, cosmic rays, and high energy photons. The energy injected by supernovae alone is proportional to SFR \citep{dekel03}, such that:
\begin{equation}
E \simeq E_{51}~\nu\ \text{SFR}~t_\mathrm{rad}
\label{eq:snenergy}
\end{equation}
Where $E_{51}$ is the total energy released by each supernova (in units of $\sim$10$^{51}$~ergs), $\nu$ is the number of supernovae per solar mass formed ($\sim$0.02/M$_\odot$), and t$_\text{rad}$ is the cooling time for an individual supernova. The energy deposition ($\dot{E}$) from supernovae and stellar winds is found from Starburst99 to be proportional to the SFR \citep{claus99, murray05}  such that
\begin{equation}
\dot{E} = 3.0 \times 10^{41} ~\text{ergs s$^{-1}$} ~\epsilon~\left(\frac{\text{SFR}}{\text{M$_\odot$ yr$^{-1}$}}\right)
\label{eq:snenergydep}
\end{equation}
Where $\epsilon$ is the thermalization efficiency factor, and is the fraction of the supernovae energy transfered into the kinetic energy of the outflow. These equations describe the total energy, and the rate that the energy is deposited into the ISM through, supernovae and stellar winds.

The combined energy from many massive stars creates an energy-conserving Sedov-Taylor blast wave \citep{taylor, sedov, mckee, weaver}. As the blast wave cools the hot gas at the shock front collapses into a thin, dense shell. This phase is called the snowplow phase, and will remain energy conserving as long as the hot, interior gas does not radiate away all of the energy \citep{lamers}. In a clumpy, inhomogenous medium, the velocity of the shell is given as \citep{mckee}:
\begin{equation}
\text{v}_\text{out} \propto E^{0.15} \propto \text{SFR}^{0.15}
\label{eq:blastwave}
\end{equation}
The SFR dependence comes from the dependence of the energy deposited by supernovae into the ISM (\autoref{eq:snenergy}). This theoretical scaling of \vout and SFR agrees within 1$\sigma$ of the relation found in \autoref{eq:scaling}. The observed \voutp-SFR relation matches a propagating shock front model.

If the star formation energy drives the outflow, then the kinetic energy deposition must be equal to the energy deposition from star formation \citep{dekel03,murray05}. From \autoref{eq:snenergydep} the kinetic energy is found to be:
\begin{equation}
\dot{E} = 1/2 ~\dot{\text{M}}_o ~\text{v}^2_\text{out} = 3.0 \times 10^{41}  \text{ergs s$^{-1}$} ~\epsilon\left(\frac{\text{SFR}}{1~M_\odot \text{yr}^{-1}}\right)
\label{eq:snkin}
\end{equation}
If we use the definition of $\eta$ ($\eta$ = \moutp/SFR) we get the relation:
\begin{equation}
\eta = \epsilon \left(\frac{\text{v}_\text{out}}{975~\text{km s$^{-1}$}}\right)^{-2}
\end{equation}
Using the median \vcen value (113~\kmsp), and an $\epsilon$ value of 0.007 (i.e. 0.7\% of the energy is converted into the kinetic energy of the warm outflow) we find an expected $\eta$ of 0.47, coincident with our measured median value. This shows that most of the energy released from supernovae is lost through radiation, adiabatic expansion, or other in other phases of the outflow. This value of $\epsilon$ is lower than the typical assumed value of 1-10\%  \citep{thornton, efstathiou}, but is consistent with recent hydrodynamical simulations \citep{simpson14}. Using energy conservation (\autoref{eq:snkin}) and the theoretical scaling of the \vout as SFR$^{0.15}$ (\autoref{eq:blastwave}), the theoretical scaling of the mass outflow rate can can be found as:
\begin{equation}
\dot{M}_o \propto \text{SFR}^{0.7}
\end{equation}
This is in 2.4$\sigma$ (1.3$\sigma$) agreement with the relations with (and without) the inclusion of SBS1415+437, and agrees within 1$\sigma$ of the multivariate fit of \mout (\autoref{eq:multimout}). The errors for this relation are quite large, and SBS1415+437 heavily impacts the fit, but the agreement is statistically consistent. The conservation of energy imparted by supernovae explains both the velocity of the outflows (\voutp) and the total amount of mass ejected in the outflow (\mout and $\eta$).

We use the supernovae energy to explain the outflow, however, this is assuredly not the only mechanism that drives outflows. Radiation pressure from photons, and cosmic rays impart momentum into the ISM that drive dense gas outward \citep{murray05, everret}. \citet{hopkins12b} present simulations where momentum deposition from these processes is essential to drive the outflows of high redshift, rapidly star-forming galaxies, similar to the mergers in this sample. It is possible that momentum deposition is the dominate driving mechanism for the mergers, and the mergers that lie off the supernovae relation are driven by radiation pressure instead of energy from supernovae. However, we would need a larger sample of merging galaxies to conclusively make this distinction. 

In summary, star formation processes impart energy into the ISM, at a rate that is proportional to the SFR, and drive a shock wave into the surrounding ISM. The shock wave sets the \vout of the gas, while energy conservation determines the mass outflow rate of the warm gas (\mout and $\eta$). These dependences explain the observed scaling relations between \mstarp, SFR, \vout and \moutp.

\subsection{Where Does the Gas Go?}
\label{escape}

\begin{figure*}
\includegraphics[width = 1.0\textwidth]{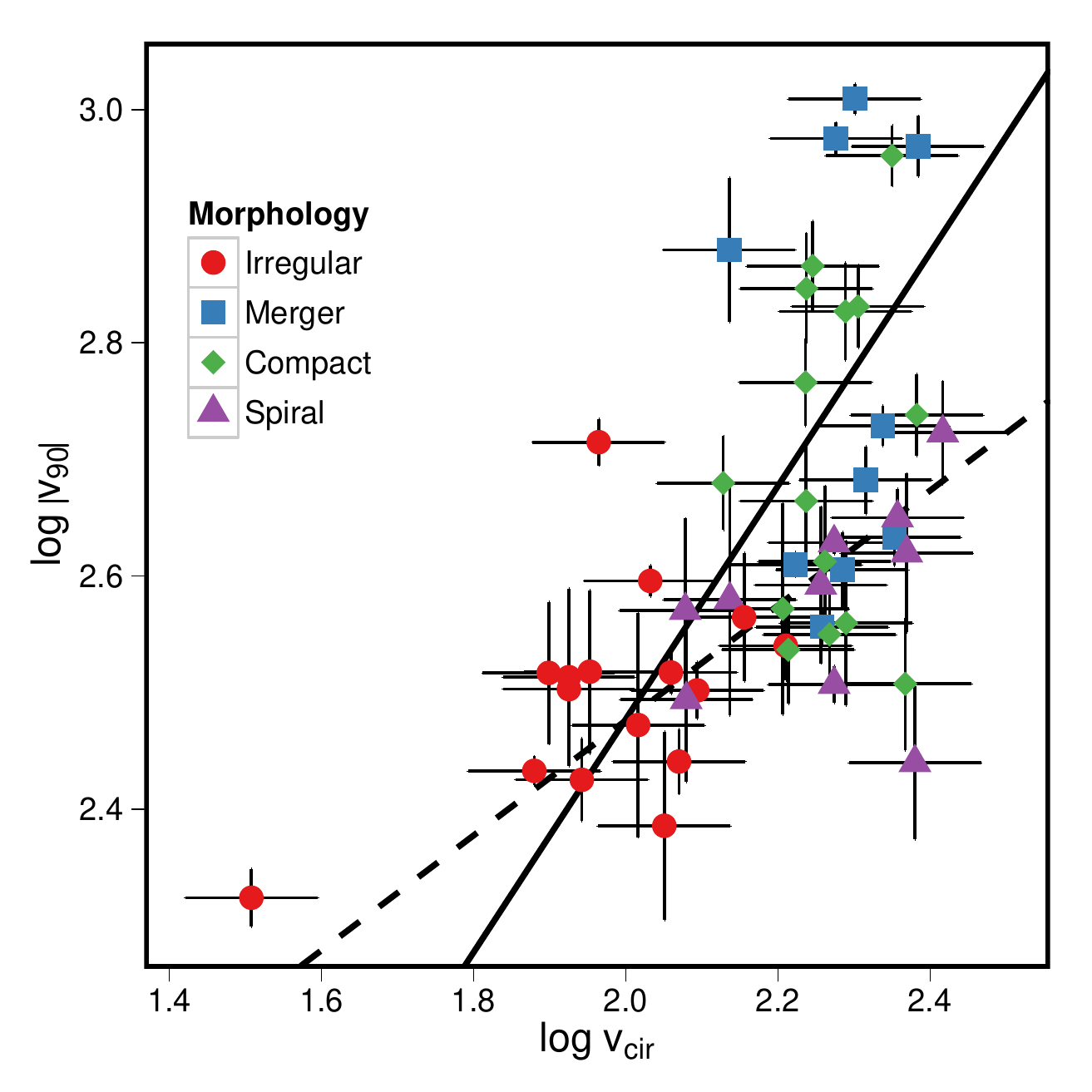}
\caption{Logarithmic relationship between \vn and \vcirc (measured in \kmsp), as calculated from the Tully-Fisher relationship. The best-fit to the data is shown as the dotted line, while the line corresponding to v$_\text{esc}$~$\simeq$~3v$_{\text{circ}}$ is shown by the solid line. Points above the solid line have outflows with material traveling faster than the escape velocity of the galaxy. The intersection point of the best-fit line of the data and the escape velocity is at log(\mstarp)~=~9.5$~\pm~0.2$~M$_\odot$ (v$_\text{circ}$~=~100~$\pm$~12~\kmsp). The low mass galaxies lie above the trend line, and are generating outflows that pollute the IGM. Additionally, there is a cluster of high mass galaxies with outflows that escape the potential well, which are either morphologically classified as mergers or are compact (see the legend for morphological classification).}
\label{fig:vesc}
\end{figure*}

The outflowing gas has two possible final locations: inside or outside the gravitational potential well of the galaxy. Here we explore how the scaling relations (\autoref{eq:scaling}) determine the final location of the gas. We first characterize the halo potential by the circular velocity of the galaxy (\vcircp) using the observed stellar mass Tully-Fisher relation (\autoref{eq:obstf}). We then find a 4$\sigma$ relationship between \vn and \vcircp, when excluding SBS1415+437, of:
\begin{equation}
\text{v}_{90} =\left(296 \pm 67~\text{\kms}\right) \left(\frac{\text{v}_{\text{circ}}}{100~\text{km s$^{-1}$}}\right)^{0.49 \pm 0.05}
\label{eq:vne}
\end{equation}
By neglecting drag forces, and assuming a singular isothermal sphere, the escape velocity at a given radius, R, is given by v$^2_\text{esc}$ = 2v$^2_\text{circ}$ ln(1+r$_s$/R), where r$_s$ is the maximum radius of the potential. Following \citet{heckman2000}, we assume that r$_s \gg R$, and simplify the relation to:
\begin{equation}
 \text{v}_\text{esc} \simeq 3 ~ \text{v}_\text{circ}  
\label{eq:vesc}
\end{equation}
Setting \vn$~=~$\vesc (\autoref{eq:vne} equal to \autoref{eq:vesc}) we find a critical log stellar mass (M$_{\ast, c}$) of 9.5$~\pm~0.2$ (\vcirc = 100~\kmsp; see \autoref{fig:vesc}). Galaxies with masses below M$_{\ast, c}$ will lose {\it some} gas to the IGM through an outflow. Therefore there are two galaxy populations: galaxies with stellar masses less than M$_{\ast, c}$ that lose some gas to the IGM through an outflow, and galaxies with masses greater M$_{\ast, c}$ that retain all of the gas accreted onto them. This gas will eventually be completely converted into stars, accreted onto a super-massive black hole, suspended in the circumgalactic medium, or lost through active galactic nuclei feedback. The low-mass galaxies will buildup stellar mass slower than the high-mass counterparts because outflows remove a fraction of the available gas reservoir for future generations of stars.

\begin{figure}[!h]
\includegraphics[width = .45\textwidth]{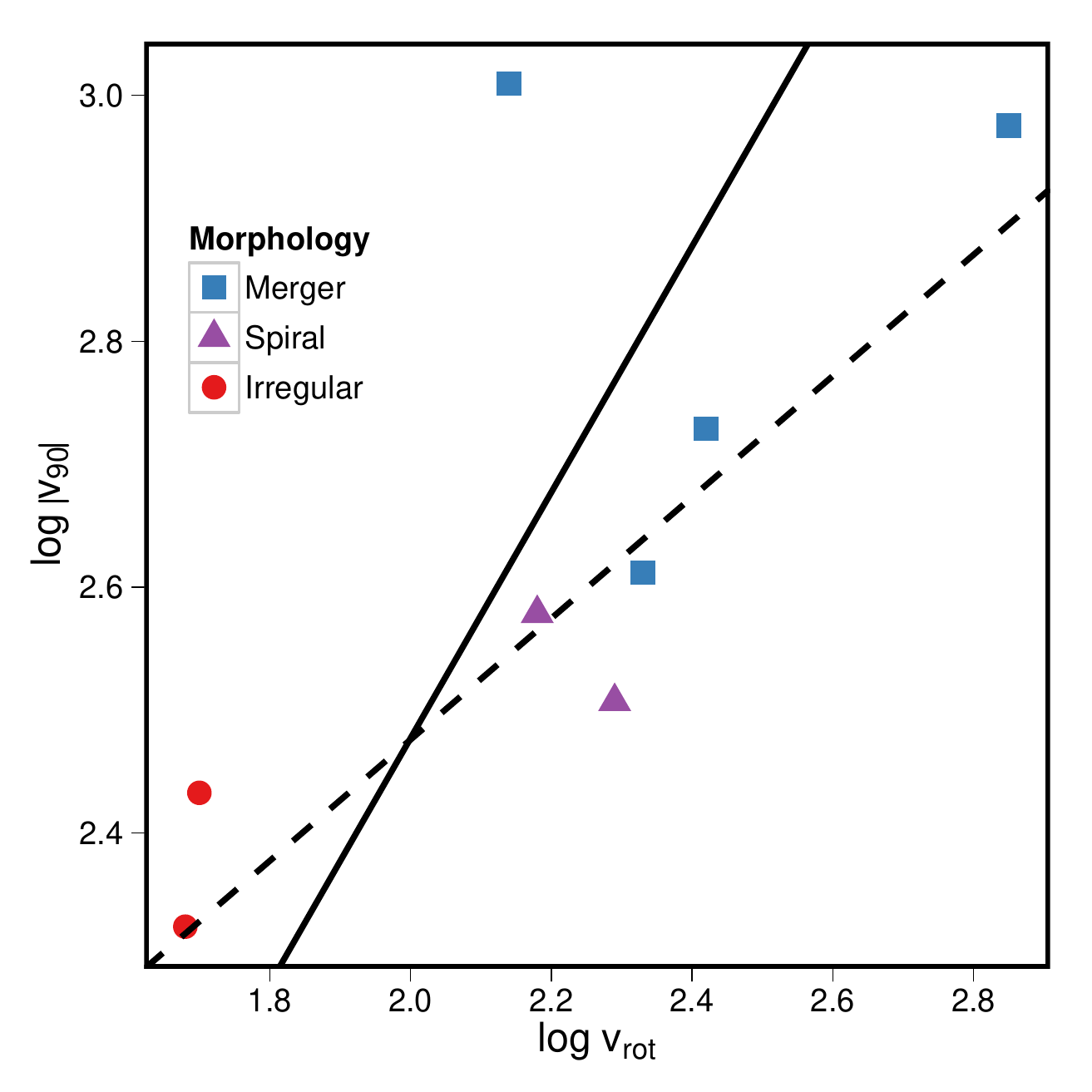}
\caption{Logarithmic relationship between \vn (in \kmsp) and the rotation velocity (v$_\text{rot}$; in \kmsp). The rotation velocity is found from the velocity at 20\% of the peak HI emission, and is only available from NED for 8 of the 51 galaxies. The lines are the same as \autoref{fig:vesc}, showing that the Tully-Fisher relation does an adequate job of estimating \vcircp. The best-fit relation to these data is \vnp~=~(38 $\pm$ 22.4 \kmsp)~v$_\text{circ}^{0.46 \pm 0.12}$, and is statistically consistent with the trend found when computing \vcirc from the Tully-Fisher relation.}
\label{fig:vrot}
\end{figure}

Since the Tully-Fisher relation is a statistical relation, we want to ensure that using the Tully-Fisher relation to estimate \vcirc  does not introduce, or bias, the trends. Unfortunately, NED has only eight galaxies with reliable HI line widths at 20\% of the emission peak.  We correct these eight widths using the standard corrections from \citet{tully85}, and then correct for the observed inclinations, although the inclination corrections may not accurately apply to the irregular and morphologically distorted galaxies. \autoref{fig:vrot} shows that we find the scaling relations between \vcirc and \vn to agree within 1$\sigma$ whether  \vcirc is calculated from the Tully-Fisher relation, or from the measured HI line widths  (the observed HI relation is: \vnp~=~[38 $\pm$ 22.4 \kmsp]~v$_\text{circ}^{0.46 \pm 0.12}$, see \autoref{fig:vrot}). 

We can also find the \mstar at which the galaxy loses the bulk of the mass in the galactic outflow. By setting \vcen$~=~$\vesc, we find that the critical mass to lose most of the gas in the outflow to be log(\mstarp) = 4.1. Very low mass systems buildup stellar mass slower than their high-mass counterparts, because outflows remove a large fraction of the available gas supply. The decelerated buildup of low-mass galaxies will act to steepen the low end of the luminosity and mass function \citep{white1978, dekel86, benson03, leja}.

Simple analytical arguments can be used to check the rough location of this critical mass. Following \citet{dekel86} and \citet{dekel03}, the value of M$_{\ast, c}$ can be estimated by assuming that the energy imparted by supernovae (\autoref{eq:snenergy}) unbinds the gas from the galaxy, or the energy from supernovae launches an outflow at the escape velocity. The authors find the maximum mass by assuming that the galaxy produces a large fraction of the total stellar mass in a single starburst, and the energy produced by this starburst removes the remaining gas from the system. \citet{dekel03} find this critical \vcirc to be the same value as found here, 100~\kmsp.

While there is a significant trend between \vn and \vcircp, there are eleven high mass galaxies with \vn~>~\vescp. From \autoref{fig:vesc} we see that these galaxies are either classified as mergers, or have a compact morphology. From the basis of their selection, in \autoref{veltrend} we argue that these high velocity, compact galaxies are most likely undergoing mergers. The only way to produce these compact objects with such a large SFR is to efficiently drive a large amount of gas into a compact area, with mergers and interactions the most likely physical mechanism to do this. During a merger the gas surface density of the galaxy rapidly increases, elevating the SFR, and driving a faster outflow (due to the outflow scaling relations). Since the gas surface densities are expected to be much higher in these cases, radiation pressure may play a large role in launching the gas (as discussed above). Gas is rapidly converted into stars, or ejected from the potential well, and the galaxy quickly settles back onto the SFR-\vout relation. This process must be faster than the dynamical time because there are many galaxies in \autoref{fig:vesc} that have merger signatures, yet are still on the \vnp-\vcirc relationship. The timescale (the dilution time) to return back onto the relation is the time to either convert the gas into stars at the SFR, or to eject the mass through an outflow. Quantitatively the dilution timescale is given as \citep{finlator08, dave12}:
\begin{equation}
t_\text{dil} = \frac{M_{gas} - M_{out}}{\text{SFR}} = \frac{t_\text{dep}}{1+\eta}
\end{equation}
NGC~7552 is an interacting galaxy with \mstar > M$_{\ast, c}$ and $\eta = 2.6$. If we assume a typical  t$_\text{dep}$ of roughly 1.5~Gyr \citep{genzel}, the galaxy will have a dilution timescale of 280~Myr, much shorter than the typical dynamical scale of 2~Gyr. Relative to the dynamics, the star formation and outflow will rapidly consume the gas reservoir, and the galaxy will settle back onto the main-sequence \voutp-\vcirc relation. 

The large scatter in these relations is likely due to stochastic mergers and accretion. In a merger the SFR is rapidly increased, which in turn increases the \vout of the galaxy. Additionally, if we assumed that the SFR depends on the accretion rate \citep{keres05, finlator08, dave12}, then the highly stochastic nature of the accretion rate  \citep{mcbride09, dekel09,fakhouri10} adds scatter to the SFR, and simultaneously to the relations, creating a relatively thick locus in the scaling relations. These mechanisms have been observed in simulations of galaxies \citep{hopkins12b}, and have been used to explain the scatter of the mass-metallicity relation \citep{finlator08}. Importantly, even though some of the galaxies fall off the scaling relations, most (78\%) of the galaxies reside along the sequence of \autoref{eq:scaling}. 

The M$_{\ast, c}$ divides the \vnp-\vcirc plane into two populations of galaxies: 1) with log(\mstarp) < 9.5, and 2) with log(\mstarp) > 9.5. Galaxies with log(\mstarp) < 9.5$~\pm~0.2$ lose gas to the IGM, while higher mass galaxies retain all of their gas. The outlying galaxies with log(\mstarp) > 9.5 and \vout > \vesc are most likely undergoing major mergers that temporarily boost the outflow velocity. We now explore how these two regimes can fundamentally shape the mass-metallicity relationship.

\subsection{The Impact of Outflows on the Mass-Metallicity Relation}
\label{metallicity}
The presence of M$_{\ast, c}$ implies that at some mass a galaxy will retain all of the gas ever accreted onto the galaxy. Since metals are a part of the total gas content, this also implies that at some mass a galaxy will retain all of the metals ever produced in the galaxy. Star formation will build up the metal content at the supernovae yield (y), while accretion of pristine gas will dilute the relative metallicity. Meanwhile, low-mass galaxies will lose gas and metals to the surrounding IGM through a galactic outflow. Assuming that galaxies reside in approximate equilibrium between star formation, accretion, and outflows, \citet{finlator08} and \citet{dave12} find a mass-metallicity slope which scales with $\eta$ as:
\begin{equation}
Z  = \frac{y}{\left(1+\eta_e\right)}\frac{1}{1-\alpha}
\label{eq:feedbackeq}
\end{equation}
Where $\alpha$ is the metallicity of the accreted IGM gas, and $\eta_e$ is the mass-loading factor of gas in the outflow that \textit{escapes} the galaxy, while the $\eta$ that we measure is the mass-loading factor of all the gas in the outflow. We decompose $\eta_e$ into two components: one describing the total mass-loading of the outflow (the measured $\eta$, with the functional form given in \autoref{tab:cor} and \autoref{eq:scaling}), and a second term describing the fraction of the outflowing gas that escapes the galaxy ($f(M_\ast)$). $f(M_\ast)$ is a complicated function that depends on the density, velocity, and geometry of the assumed outflow, and how these factors relate to the escape velocity. From \autoref{escape} we find that $f(M_\ast) = 0$ above log(\mstarp) = 9.5. It is beyond the scope of this work to solve for this function, but due to the very shallow scaling of the observed velocity structure (\autoref{eq:scaling}), we make the crude assumption that f(\mstarp) is nearly constant, or at most scales as \vn does (as M$_\ast^{0.11}$). Putting this together, gives the metallicity of a galaxy by
\begin{equation}
Z \propto \frac{y}{\left(1 + C_\eta f(M_\ast) M_\ast^{-0.58 \pm 0.1}\right)}\left(\frac{1}{1-\alpha}\right)
\end{equation}
When galaxies have \mstar < M$_{\ast, c}$, and $\eta_e > 1$ (appropriate for the low mass regime, see \autoref{fig:eff}), we find the scaling of the mass-metallicity relation to be
\begin{equation}
Z \propto \frac{M_\ast^{0.58 \pm 0.1}}{1-\alpha}
\end{equation}
While the mass-metallicity relation is Z $\propto$ y/(1-$\alpha$) for the galaxies above M$_{\ast, c}$. In log-space, this produces a mass-metallicity relation that scales linearly with a slope of $0.58 \pm 0.1$ for galaxies below M$_{\ast, c}$, and approaches the supernovae yield, modulo depletion from accretion, for high-mass galaxies.

The shape of the described mass-metallicity relation can be compared to the observed mass-metallicity relation. The observed mass-metallicity relationship has two regimes: 1) a low-mass regime with a smoothly increasing slope of $0.5 \pm 0.2$ and 2) a high-mass regime, where the metallicity is approximately equal to the supernova yield \citep{tremonti04, erb06, andrews13, zahid}. The  mass-metallicity relation has been highly debated because the exact scaling depends on the calibration system used \citep{kewley08}, and the parameterization of the relation \citep{moustakas, zahid}. These factors produce discrepancies between various measured relations. \citet{zahid} parameterize the mass-metallicity relation as 
\begin{equation}
\log{Z}  = Z_0 + \log{\left(1 - \exp{\left[-\left(\frac{M_\ast}{M_0}\right)^\gamma\right]}\right)}
\end{equation}
Where M$_0$ is the turnover mass, above which the metallicity is equal to Z$_0$, and $\gamma$ is the slope of the low-mass relation. \citet{zahid} use SDSS, DEEP2, and COSMOS data, along with the strong line method, to find the values of $\gamma$ = 0.51 and log(M$_0$) = 9.1 + 2.6log(1+z), in agreement with the shape found here. The increase of the turnover mass with increasing redshift is consistent with SFR increasing with increasing redshift  \citep{madau14}, and accordingly launching more energetic outflows capable of escaping deeper potential wells. This means that the turnover mass should scale similarly to the star formation history of the universe. This shows that the outflow scaling relations of \autoref{eq:scaling} quantitatively, and qualitatively, produce the observed mass-metallicity relation.

\section{CONCLUSION}
\label{conclusion}

Here, we present the first results from a COS-HST Archival study of 51 nearby star-forming galaxies. Using \siiip as a tracer of the warm outflowing gas, we measure the outflow velocity (\voutp), column density (\nsp), and mass outflow rate (\moutp) from the COS UV spectra. Stellar masses (\mstarp), star formation rates (SFRs), star formation rate surface densities (\sfrsdp), inclinations, and morphologies are measured from the public data archives of \wisep, \galexp, and the SDSS. Combined, this sample covers over four orders of magnitude in SFR and \mstarp, and samples a wide range of galactic morphologies.

By combining the outflow properties and host galaxy properties, we measure the scaling relations between outflows and host galaxies. We find:
\begin{itemize}
\item The outflow velocity (both centroid velocity and the velocity at 90\% of the continuum) varies significantly, yet shallowly, with both SFR and \mstarp. We find scalings of SFR$^{0.08\rightarrow0.14}$ and M$_\ast^{0.08\rightarrow0.12}$ for \vnp$\rightarrow$\vcen (see \autoref{eq:scaling} and \autoref{eq:scaling1}).

\item Using the observed Tully-Fisher relation we find the outflow velocity to scale with the circular velocity as v$_\text{circ}^{0.49}$.

\item Some galaxies with merger and interaction signatures drive faster outflows than their normal counterparts, while other merging and interacting galaxies drive outflows that lie along the mean relations.

\item We find statistically significant relations between both \mout and the mass-loading efficiencies ($\eta$~=~\moutp/SFR) with SFR, \mstarp, and \vcirc. To the nearest fraction, these relations are \moutp~$\propto$~M$_\ast^{1/2}$, SFR$^{1/2}$, and $\eta$~$\propto$~M$_\ast^{-1/2}$, v$_\text{circ}^{-5/2}$ (see \autoref{eq:scaling} and \autoref{eq:scalingcirc}).

\item We find a statistically significant multivariate relationship between \moutp, \mstarp, and SFR with a scaling of \moutp~$\propto$~SFR$^{3/4} $M$_\ast^{-1/3}$ (\autoref{eq:multimout}). We do not find other statistically significant multivariate relations.

\item We do not find statistically significant correlations between any outflow property and \sfrsdp, inclination, or sSFR (SFR/\mstarp).
\end{itemize}
From these results we discuss the impact of outflows on galaxy evolution:
\begin{itemize}
\item The velocity and energetics suggest that feedback from supernovae drives a shock front into the surrounding ISM, and the outflow moves outward at a velocity that depends on the SFR. The observations require only 0.7\% of the total supernovae energy to be converted into the kinetic energy of the warm outflow phase. Some merging galaxies may have a different driving mechanism.

\item We determine a critical mass (M$_{\ast, c}$) below which a galaxy will lose {\it some} gas to the intergalactic medium through an outflow. We find this mass to be log(M$_{\ast, c}$)~=~9.5$~\pm~0.2$.

\item We find that the critical mass splits the galaxies into two populations: high mass galaxies that retain all of the accreted gas, and low mass galaxies that lose a fraction of the accreted gas.

\item We discuss how these two galaxy populations, and the outflow scaling relations, qualitatively, and quantitatively, describes the mass-metallicity relation.

\end{itemize} 
In future work we plan to further this investigation by extending the analysis to different elements and species to better understand the ionization structure of galactic outflows. Further, there is a strong need for more UV spectroscopic observations of high-mass, \lq\lq{}normal\rq\rq{} star-forming galaxies (i.e. ones that lie on the star-forming main-sequence, as defined by the SDSS), as well as galaxies undergoing mergers. This will help break the strong linear trend between \mstar and SFR, observed here, and may further aid in understanding the degeneracy between the \mstar and SFR in the outflow scaling relations.

\section*{Acknowledgements}
We thank Bart Waaker for the extensive help in extraction and reduction of the COS data. 

Support for program 13239 was provided by NASA through a grant from the Space Telescope Science Institute, which is operated by the Association of Universities for Research in Astronomy, Inc., under NASA contract NAS 5-26555. Some of the data presented in this paper were obtained from the Mikulski Archive for Space Telescopes (MAST). STScI is operated by the Association of Universities for Research in Astronomy, Inc., under NASA contract NAS 5-26555.  

This research has made use of the NASA/IPAC Extragalactic Database (NED), which is operated by the Jet Propulsion Laboratory, California Institute of Technology, under contract with the National Aeronautics and Space Administration. This research has made use of the NASA/ IPAC Infrared Science Archive, which is operated by the Jet Propulsion Laboratory, California Institute of Technology, under contract with the National Aeronautics and Space Administration. This publication makes use of data products from the Wide-field Infrared Survey Explorer, which is a joint project of the University of California, Los Angeles, and the Jet Propulsion Laboratory/California Institute of Technology, funded by the National Aeronautics and Space Administration.

Funding for the Sloan Digital Sky Survey (SDSS) has been provided by the Alfred P. Sloan Foundation, the Participating Institutions, the National Aeronautics and Space Administration, the National Science Foundation, the U.S. Department of Energy, the Japanese Monbukagakusho, and the Max Planck Society. The SDSS Web site is http://www.sdss.org/. The SDSS is managed by the Astrophysical Research Consortium (ARC) for the Participating Institutions. The Participating Institutions are The University of Chicago, Fermilab, the Institute for Advanced Study, the Japan Participation Group, The Johns Hopkins University, Los Alamos National Laboratory, the Max-Planck-Institute for Astronomy (MPIA), the Max-Planck-Institute for Astrophysics (MPA), New Mexico State University, University of Pittsburgh, Princeton University, the United States Naval Observatory, and the University of Washington.

\appendix
\label{tables}
Here we provide tables of the galaxy data (\autoref{tab:target}), the derived host galaxy properties (\autoref{tab:gal}), and the derived outflow properties (\autoref{tab:outflow}). 
\begin{deluxetable}{cccccccc}
\tablewidth{0pt}
\tablecaption{Target Data}
\tablehead{
\colhead{Galaxy Name} &
\colhead{Project ID} & 
\colhead{RA} & 
\colhead{DEC} & 
\colhead{z} & 
\colhead{D} &
\colhead{Morphology} &
\colhead{12+log(O/H)} 
\\ 
\colhead{} &
\colhead{} &
\colhead{hh mm ss} &
\colhead{hh mm ss} &
\colhead{} &
\colhead{(Mpc)} &
\colhead{} &
\colhead{} 
}
\startdata
IRAS 08339+6517  &  12173  &08 38 23.150 &+65 07 15.40  &0.01911  &     83  &M   &8.77  \\
NGC 3256         &  12173  &10 27 51.340 &-43 54 12.40  &0.00935  &     40  &M   &8.78  \\
NGC 6090         &  12173  &16 11 40.200 &+52 27 23.75  &0.02930  &    128  &M   &8.78  \\
NGC 7552         &  12173  &23 16 10.790 &-42 35 05.50  &0.00537  &     23  &M   &8.77  \\
Haro 11          &  13017  &00 36 51.820 &-33 33 17.00  &0.02060  &     89  &M   &8.77  \\
J0055-0021      &  11727  &00 55 27.470 &-00 21 48.55  &0.16740  &    806  &C   &8.78  \\
J0150+1260      &  11727  &01 50 28.410 &+13 08 58.40  &0.14670  &    697  &M   &8.78  \\
J0213+1259      &  11727  &02 13 48.530 &+12 59 51.40  &0.21900  &   1089  &S   &8.78  \\
J0938+5428      &  11727  &09 38 13.490 &+54 28 25.09  &0.10210  &    471  &M   &8.76  \\
J2103-0728      &  11727  &21 03 58.740 &-07 28 02.48  &0.13660  &    645  & S  &8.79  \\
M83\_1           &  11579  &13 37 00.516 &-29 52 00.48  &0.00171  &      4  &S   &8.77  \\
M83\_2           &  11579  &13 37 00.516 &-29 52 00.48  &0.00171  &      4  &S   &8.77  \\
NGC 4449         &  11579  &12 28 10.816 &+44 05 42.95  &0.00069  &      3  &I   &8.76  \\
NGC 5253       &  11579  &13 39 55.976 &-31 38 27.01  &0.00136  &      3  &I   &8.52  \\
SBS 1415+437     &  11579  &14 17 01.406 &+43 30 04.75  &0.00203  &     13  &I   &7.71  \\
KISSR 1084       &  11522  &16 49 05.260 &+29 45 31.60  &0.03208  &    140  &S   &8.77  \\
KISSR 1578       &  11522  &13 28 44.050 &+43 55 50.50  &0.02795  &    122  &I   &8.69  \\
KISSR 1637       &  11522  &13 42 00.005 &+42 45 36.15  &0.03471  &    152  &I   &8.69  \\
KISSR 2021       &  11522  &15 45 52.780 &+44 15 47.60  &0.03996  &    176  &S   &8.70  \\
KISSR 218        &  11522  &13 09 16.139 &+29 22 02.61  &0.02093  &     91  &S   &8.73  \\
KISSR 242        &  11522  &13 16 03.900 &+29 22 53.80  &0.03785  &    166  &I   &8.71  \\
KISSR 108        &  12027  &12 43 55.300 &+29 22 10.70  &0.02362  &    103  &I   &8.65  \\
KISSR 178        &  12027  &13 01 41.534 &+29 22 52.24  &0.05740  &    256  &I   &8.74  \\
KISSR 182        &  12027  &13 02 25.650 &+28 51 29.10  &0.02236  &     97  &I   &8.69  \\
KISSR 1942       &  12027  &15 22 48.480 &+43 46 24.50  &0.03940  &    173  &I   &8.83  \\ 
KISSR 271        &  12027  &13 21 40.800 &+28 52 59.10  &0.02264  &     98  &I   &8.74  \\
KISSR 326        &  12027  &13 39 16.100 &+28 52 25.00  &0.04536  &    201  &S   &8.70  \\
KISSR 40         &  12027  &12 22 23.600 &+29 26 37.80  &0.02622  &    114  &I   &8.69  \\
IRAS 09583+4714  &  12533  &10 01 30.45 &+46 59 57.74  &0.08591  &    392  &M   &8.73  \\
IRAS 16487+5447  &  12533  &16 49 47.031 &+54 42 35.41  &0.10376  &    479  &C   &8.78  \\
NGC 7714         &  12604  &17 03 42.008 &+58 13 44.39  &0.00933  &     40  &M   &8.76  \\
J0824+2806      &  13017  &08 23 54.96 &+28 06 21.67  &0.04722  &    209  &M   &8.77  \\
J1113+2930      &  13017  &11 13 23.880 &+29 30 39.32  &0.17510  &    847  &C   &8.75  \\
J1144+4012      &  13017  &11 44 22.280 &+40 12 21.19  &0.12700  &    596  &C   &8.76  \\
J1429+1653      &  13017  &14 28 56.410 &+16 53 39.32  &0.18160  &    882  &C   &8.77  \\
GP0303-0759     &  12928  &03 03 21.414 &-07 59 23.25  &0.16500  &    793  &C   &8.76  \\
GP0911+1831     &  12928  &09 11 13.345 &+18 31 08.17  &0.26220  &   1335  &C   &8.76  \\
GP1133+6514     &  12928  &11 33 03.804 &+65 13 41.38  &0.24140  &   1215  &C   &8.77  \\
GP1244+0216     &  12928  &12 44 23.372 &+02 15 40.43  &0.23950  &   1204  &C   &8.77  \\
J1416+1223      &  13017  &14 16 12.870 &+12 23 40.42  &0.12310  &    576  &C   &8.77  \\
J1525+0757      &  13017  &15 25 21.880 &+07 57 20.30  &0.07570  &    343  &C   &8.76  \\
J1429+0643      &  13017  &14 29 47.000 &+06 43 34.95  &0.17360  &    839  &C   &8.76  \\
J1112+5503      &  13017  &11 12 44.050 &+55 03 47.01  &0.13150  &    619  &C   &8.78  \\
J1025+3622      &  13017  &10 25 48.380 &+36 22 58.42  &0.12650  &    593  &C   &8.73  \\
MRK 1486         &  12583  &13 59 50.918 &+57 26 22.76  &0.03377  &    148  &I   &8.86  \\
J0907+5327      &  12583  &09 07 04.894 &+53 26 57.48  &0.02986  &    130  &I   &8.72  \\
J1250+0734      &  12583  &12 50 13.649 &+07 34 47.26  &0.03821  &    168  &S   &8.78  \\
J1307+5427      &  12583  &13 07 28.675 &+54 26 49.48  &0.03252  &    142  &I   &8.70  \\
J1315+6207      &  12583  &13 15 35.160 &+62 07 28.11  &0.03079  &    135  &M   &8.78  \\
J1403+0628      &  12583  &14 03 47.237 &+06 28 12.39  &0.08433  &    384  &S   &8.78  \\
GP1054+5238     &  12928  &10 53 30.823 &+52 37 52.87  &0.25260  &   1279  &C   &8.77  \\
\enddata
\label{tab:target}
\tablecomments{Table of the target information. Column 1 gives the galaxy name. Column 2 is the HST project identification number. Columns 3 and 4 are the RA and Dec of the COS aperture. Column 5 is the redshift used, while column 6 is the distance (in Mpc) calculated from either the Hubble flow, or from redshift independent distances from NED. Column 7 is the morphology of the galaxy with the following key: M stands for mergers, I stands from irregulars, S stands for spirals, and c stands for compact galaxies. The best-fit metallicity (in units of 12+log(O/H)) used to create the Starburst99 theoretical continuum is given in column 8. This metallicity is derived from the mass-metallicity relation of either \citet{tremonti04} or \citet{andrews13}.}
\end{deluxetable}

\begin{deluxetable}{cccccccc}
\tabletypesize{\small}
\tablewidth{0pt}
\tablecaption{Galaxy Properties}
\tablehead{
\colhead{Galaxy Name} &
\colhead{M$_\ast$} & 
\colhead{SFR} & 
\colhead{$\Sigma_\text{{SFR}}$} & 
\colhead{$i$} & 
\colhead{R} &
\colhead{log(Age)} &
\colhead{SF Law} \\ 
\colhead{} &
\colhead{(log(M$_\ast$/M$_\odot$))} & 
\colhead{(M$_\odot$ yr$^{-1}$)} & 
\colhead{(M$_\odot$ yr$^{-1}$ kpc$^{-2}$)} & 
\colhead{($^\circ$)} & 
\colhead{(kpc)} &
\colhead{(log(yr$^{-1}$))} &
\colhead{}
}
\startdata
IRAS 08339+6517  &10.74  & 13.59 & 6.42  &     29  &  4.4  &  7.1   &I  \\
NGC 3256         &11.23  & 50.83 & 1.17  &     48  &  7.0  &  6.6   &I  \\
NGC 6090         &11.02  & 25.15 & 3.28  &     29  &  5.5  &  6.6   &I  \\
NGC 7552         &10.85  & 13.37 & 3.14  &     56  &  4.4  &  6.6   &I  \\
Haro 11          &10.66  & 26.45 &14.31  &     56  &  5.6  &  7.1   &I  \\
J0055-0021      &11.22  & 71.15 & 1.76  &     54  &  7.9  &  8.0   &C  \\
J0150+1260      &11.09  & 60.40 & 1.34  &     53  &  7.4  &  8.0   &C  \\
J0213+1259      &11.21  & 66.12 & 0.86  &      0  &  7.7  &  7.1   &I  \\
J0938+5428      &10.44  & 20.28 & 1.08  &     61  &  5.1  &  7.1   &I  \\
J2103-0728      &11.37  &136.82 & 3.71  &     50  &  9.9  &  6.6   &I  \\
M 83\_1           &10.73  &  3.65 & 5.64  &     34  &  2.8  &  6.6   &I  \\
M 83\_2           &10.73  &  3.65 &17.93  &     34  &  2.8  &  6.6   &I  \\
NGC 4449         & 9.34  &  0.24 & 2.61  &     69  &  1.0  &  7.1   &I  \\
NGC 5253      & 8.96  &  0.48 & 2.73  &     56  &  1.3  &  7.3   &I  \\
SBS 1415+437     & 7.29  &  0.02 & 0.75  &     53  &  0.5  &  6.9   &I  \\
KISSR 1084       &10.65  &  3.00 & 0.23  &      0  &  2.6  &  6.6   &I  \\
KISSR 1578       & 9.77  &  3.72 & 1.39  &     32  &  2.8  &  8.0   &C  \\
KISSR 1637       & 9.73  &  0.82 & 0.06  &     58  &  1.6  &  7.1   &I  \\
KISSR 2021       & 9.85  &  1.32 & 0.05  &     53  &  1.9  &  7.1   &I  \\
KISSR 218        &10.11  &  0.99 & 0.15  &     48  &  1.7  &  7.1   &I  \\
KISSR 242        & 9.92  &  5.12 & 1.37  &     71  &  3.1  &  8.0   &C  \\
KISSR 108        & 9.05  &  0.23 & 0.15  &     55  &  1.0  &  8.0   &C  \\
KISSR 178        &10.20  &  2.55 & 0.18  &     68  &  2.4  &  6.6   &I  \\
KISSR 182        & 9.16  &  0.28 & 0.16  &     67  &  1.1  &  7.1   &I  \\
KISSR 1942       & 9.57  &  0.75 & 0.02  &     63  &  1.6  &  7.1   &I  \\
KISSR 271        & 9.29  &  0.18 & 0.04  &     43  &  1.0  &  7.1   &I  \\
KISSR 326        & 9.86  &  1.43 & 0.01  &      0  &  2.0  &  8.0   &C  \\
KISSR 40         & 9.16  &  0.28 & 0.04  &     32  &  1.1  &  8.0   &C  \\
IRAS 09583+4714  &10.11  &  5.84 & 0.46  &     57  &  3.3  &  6.6   &I  \\
IRAS 16487+5447  &11.15  & 88.89 & 0.61  &     35  &  8.5  &  6.6   &I  \\
NGC 7714         &10.50  &  9.17 &17.35  &     35  &  3.8  &  7.1   &I  \\
J0824+2806      &10.78  & 20.73 & 1.22  &     44  &  5.1  &  8.0   &C  \\
J1113+2930      &10.43  & 11.71 & 0.14  &     46  &  4.2  &  7.1   &I  \\
J1144+4012      &10.57  & 11.82 & 0.28  &     25  &  4.2  &  7.1   &I  \\
J1429+1653      &10.70  & 28.63 & 0.43  &     57  &  5.7  &  7.1   &I  \\
GP0303-0759     &10.56  & 24.66 & 0.61  &     55  &  5.4  &  8.0   &C  \\
GP0911+1831     &10.46  & 34.70 & 0.40  &     55  &  6.1  &  8.0   &C  \\
GP1133+6514     &10.80  & 40.78 & 0.42  &     50  &  6.5  &  8.0   &C  \\
GP1244+0216     &10.80  & 71.73 & 0.89  &     41  &  7.9  &  8.0   &C  \\
J1416+1223      &10.87  & 46.60 & 1.68  &      0  &  6.8  &  8.0   &C  \\
J1525+0757      &10.60  & 12.97 & 0.92  &     52  &  4.3  &  8.0   &C  \\
J1429+0643      &10.56  & 61.07 & 1.42  &     33  &  7.5  &  8.0   &C  \\
J1112+5503      &11.07  & 58.47 & 2.16  &      0  &  7.4  &  8.0   &C  \\
J1025+3622      &10.08  & 13.03 & 0.52  &     67  &  4.3  &  8.0   &C  \\
MRK 1486         & 9.64  &  3.60 & 1.10  &     85  &  2.7  &  7.1   &I  \\
J0907+5327      & 9.24  &  0.86 & 0.19  &     77  &  1.7  &  8.0   &C  \\
J1250+0734      &11.10  & 16.71 & 0.79  &     70  &  4.7  &  7.1   &I  \\
J1307+5427      & 9.81  &  5.69 & 0.48  &     58  &  3.2  &  8.0   &C  \\
J1315+6207      &10.92  & 41.08 & 6.82  &     58  &  6.5  &  7.1   &I  \\
J1403+0628      &11.16  & 21.27 & 0.21  &     87  &  5.1  &  7.1   &I  \\
GP1054+5238     &10.67  & 46.20 & 0.60  &     22  &  6.8  &  7.1   &I  \\
\enddata
\label{tab:gal}
\tablecomments{Table of the calculated galaxy properties. Column 1 gives the galaxy name. Column 2 is the stellar mass (M$_\ast$) of the galaxy. Columns 3 is the star formation rate (SFR) of the entire galaxy. Column 4 is the star formation rate surface density ($\Sigma_\text{SFR}$) inside the COS aperture.  Column 5 is the inclination ($i$) of the galaxy. Column 6 gives the radius of the outflow (R) used to calculate the mass outflow rate. The stellar population age (log(age)) calculated from the Starburst99 theoretical spectra is given in column 7, with the star formation law given in column 8. The star formation law is given as either continuous (C) star formation, or an instantiates burst (I).}
\end{deluxetable}

\begin{deluxetable}{cccccc}
\tabletypesize{\small}
\tablewidth{0pt}
\tablecaption{Outflow Properties}
\tablehead{
\colhead{Galaxy Name} &
\colhead{v$_\text{cen}$} & 
\colhead{v$_{90}$} & 
\colhead{log(N$_\text{Si~{\sc II}}$)} & 
\colhead{$\dot{M}_\text{out}$} & 
\colhead{$\eta$} \\
\colhead{} &
\colhead{(km s$^{-1}$)} & 
\colhead{(km s$^{-1}$)} & 
\colhead{(cm$^{-2}$)} & 
\colhead{({M}$_\odot$ yr$^{-1}$)} & 
\colhead{} }
\startdata
IRAS 08339+6517  &        -386 $\pm$          11  &        -944 $\pm$          29 &16.31 $\pm$ 0.02&39.94 $\pm$ 8.17  & 2.939 $\pm$ 0.841  \\
NGC 3256         &        -371 $\pm$          18  &        -930 $\pm$          55 &15.71 $\pm$ 0.23&37.91 $\pm$15.23  & 0.746 $\pm$ 0.335  \\
NGC 6090         &        -177 $\pm$          12  &        -535 $\pm$          21 &14.94 $\pm$ 0.20& 2.80 $\pm$ 1.03  & 0.111 $\pm$ 0.047  \\
NGC 7552         &        -454 $\pm$          16  &       -1022 $\pm$          29 &15.77 $\pm$ 0.11&34.29 $\pm$ 8.94  & 2.565 $\pm$ 0.843  \\
Haro 11          &        -163 $\pm$           9  &        -360 $\pm$          13 &15.00 $\pm$ 0.13& 1.70 $\pm$ 0.48  & 0.064 $\pm$ 0.022  \\
J0055-0021      &        -124 $\pm$           6  &        -547 $\pm$          44 &15.61 $\pm$ 0.68&11.82 $\pm$12.37  & 0.166 $\pm$ 0.177  \\
J0150+1260      &        -154 $\pm$           7  &        -429 $\pm$          23 &15.98 $\pm$ 0.07&28.80 $\pm$ 6.69  & 0.477 $\pm$ 0.146  \\
J0213+1259      &        -119 $\pm$          32  &        -275 $\pm$          41 &14.56 $\pm$ 0.11& 1.07 $\pm$ 0.41  & 0.016 $\pm$ 0.007  \\
J0938+5428      &         -19 $\pm$           5  &        -346 $\pm$          23 &14.94 $\pm$ 0.62& 0.25 $\pm$ 0.25  & 0.012 $\pm$ 0.013  \\
J2103-0728      &        -165 $\pm$           9  &        -528 $\pm$          53 &15.75 $\pm$ 0.23&29.24 $\pm$11.91  & 0.214 $\pm$ 0.097  \\
M 83\_1           &         -65 $\pm$           6  &        -425 $\pm$           9 &15.49 $\pm$ 0.10& 1.73 $\pm$ 0.47  & 0.474 $\pm$ 0.160  \\
M 83\_2           &         -86 $\pm$           6  &        -321 $\pm$          11 &15.30 $\pm$ 0.01& 1.66 $\pm$ 0.36  & 0.456 $\pm$ 0.134  \\
NGC 4449         &        -111 $\pm$          16  &        -518 $\pm$          23 &14.94 $\pm$ 0.04& 0.38 $\pm$ 0.10  & 1.596 $\pm$ 0.520  \\
NGC 5253       &         -90 $\pm$           5  &        -270 $\pm$           7 &15.47 $\pm$ 0.01& 2.30 $\pm$ 0.48  & 4.820 $\pm$ 1.397  \\
SBS 1415+437     &         -54 $\pm$           6  &        -210 $\pm$          12 &15.04 $\pm$ 0.01& 1.16 $\pm$ 0.28  &52.552 $\pm$16.341  \\
KISSR 1084       &        -159 $\pm$          15  &        -390 $\pm$          60 &15.82 $\pm$ 0.29& 8.40 $\pm$ 4.10  & 2.796 $\pm$ 1.475  \\
KISSR 1578       &        -116 $\pm$           5  &        -328 $\pm$          13 &14.83 $\pm$ 0.47& 0.92 $\pm$ 0.68  & 0.247 $\pm$ 0.189  \\
KISSR 1637       &         -76 $\pm$          15  &        -243 $\pm$          44 &15.56 $\pm$ 0.19& 1.52 $\pm$ 0.62  & 1.856 $\pm$ 0.843  \\
KISSR 2021       &         -81 $\pm$          13  &        -371 $\pm$          67 &15.66 $\pm$ 0.22& 2.86 $\pm$ 1.20  & 2.173 $\pm$ 1.011  \\
KISSR 218        &         -21 $\pm$          17  &        -379 $\pm$          86 &15.13 $\pm$ 0.81& 0.18 $\pm$ 0.27  & 0.182 $\pm$ 0.272  \\
KISSR 242        &         -91 $\pm$           5  &        -317 $\pm$          18 &15.68 $\pm$ 0.02& 4.55 $\pm$ 0.96  & 0.888 $\pm$ 0.258  \\
KISSR 108        &         -63 $\pm$          15  &        -328 $\pm$          46 &15.34 $\pm$ 0.66& 0.62 $\pm$ 0.65  & 2.658 $\pm$ 2.834  \\
KISSR 178        &         -43 $\pm$          10  &        -366 $\pm$          47 &14.98 $\pm$ 0.28& 0.38 $\pm$ 0.20  & 0.149 $\pm$ 0.083  \\
KISSR 182        &         -71 $\pm$           7  &        -325 $\pm$          56 &14.83 $\pm$ 0.75& 0.18 $\pm$ 0.21  & 0.658 $\pm$ 0.774  \\
KISSR 1942       &         -38 $\pm$          27  &        -296 $\pm$          66 &15.68 $\pm$ 0.42& 0.90 $\pm$ 0.89  & 1.194 $\pm$ 1.205  \\
KISSR 271        &         -79 $\pm$          58  &        -329 $\pm$          53 &15.89 $\pm$ 0.54& 1.44 $\pm$ 1.62  & 7.922 $\pm$ 9.031  \\
KISSR 326        &        -112 $\pm$          59  &        -311 $\pm$          51 &15.74 $\pm$ 0.76& 4.38 $\pm$ 5.63  & 3.066 $\pm$ 3.984  \\
KISSR 40         &         -75 $\pm$          20  &        -318 $\pm$          48 &15.38 $\pm$ 0.24& 0.48 $\pm$ 0.24  & 1.745 $\pm$ 0.926  \\
IRAS 09583+4714  &        -198 $\pm$          23  &        -758 $\pm$         108 &15.97 $\pm$ 0.05&14.48 $\pm$ 3.55  & 2.477 $\pm$ 0.784  \\
IRAS 16487+5447  &         -28 $\pm$           8  &        -321 $\pm$          41 &15.68 $\pm$ 0.06& 3.72 $\pm$ 1.38  & 0.042 $\pm$ 0.018  \\
NGC 7714         &        -107 $\pm$           4  &        -407 $\pm$           9 &14.92 $\pm$ 0.02& 1.13 $\pm$ 0.23  & 0.124 $\pm$ 0.036  \\
J0824+2806      &         -32 $\pm$           7  &        -402 $\pm$          29 &15.83 $\pm$ 0.13& 3.75 $\pm$ 1.39  & 0.181 $\pm$ 0.076  \\
J1113+2930      &        -104 $\pm$          15  &        -373 $\pm$          78 &14.39 $\pm$ 0.47& 0.22 $\pm$ 0.17  & 0.019 $\pm$ 0.015  \\
J1144+4012      &        -236 $\pm$           9  &        -702 $\pm$          75 &15.47 $\pm$ 0.38& 9.95 $\pm$ 6.08  & 0.842 $\pm$ 0.542  \\
J1429+1653      &        -128 $\pm$           9  &        -354 $\pm$          33 &15.55 $\pm$ 0.07& 6.15 $\pm$ 1.47  & 0.215 $\pm$ 0.067  \\
GP0303-0759     &        -232 $\pm$          24  &        -461 $\pm$          52 &15.20 $\pm$ 0.24& 2.85 $\pm$ 1.22  & 0.115 $\pm$ 0.055  \\
GP0911+1831     &        -208 $\pm$          15  &        -344 $\pm$          37 &15.51 $\pm$ 0.08& 6.73 $\pm$ 1.67  & 0.194 $\pm$ 0.062  \\
GP1133+6514     &        -429 $\pm$          27  &        -671 $\pm$          65 &14.75 $\pm$ 0.12& 5.20 $\pm$ 1.42  & 0.128 $\pm$ 0.043  \\
GP1244+0216     &         -78 $\pm$           8  &        -362 $\pm$          58 &15.71 $\pm$ 0.08& 9.46 $\pm$ 2.44  & 0.132 $\pm$ 0.043  \\
J1416+1223      &        -124 $\pm$           8  &        -677 $\pm$          55 &15.93 $\pm$ 0.03&20.86 $\pm$ 4.51  & 0.448 $\pm$ 0.132  \\
J1525+0757      &        -385 $\pm$          20  &        -734 $\pm$          66 &15.30 $\pm$ 0.44& 8.26 $\pm$ 5.77  & 0.637 $\pm$ 0.463  \\
J1429+0643      &        -218 $\pm$          10  &        -583 $\pm$          50 &16.01 $\pm$ 0.10&56.32 $\pm$14.46  & 0.922 $\pm$ 0.300  \\
J1112+5503      &        -387 $\pm$          16  &        -913 $\pm$          53 &15.87 $\pm$ 0.37&51.48 $\pm$30.81  & 0.880 $\pm$ 0.556  \\
J1025+3622      &        -188 $\pm$           6  &        -478 $\pm$          43 &15.94 $\pm$ 0.05&20.68 $\pm$ 4.47  & 1.586 $\pm$ 0.467  \\
MRK 1486         &        -112 $\pm$           8  &        -394 $\pm$          12 &14.69 $\pm$ 0.59& 0.30 $\pm$ 0.28  & 0.084 $\pm$ 0.079  \\
J0907+5327      &         -63 $\pm$           6  &        -266 $\pm$          21 &15.53 $\pm$ 0.06& 1.41 $\pm$ 0.35  & 1.640 $\pm$ 0.518  \\
J1250+0734      &        -193 $\pm$           6  &        -446 $\pm$          24 &15.19 $\pm$ 0.64& 4.47 $\pm$ 4.44  & 0.268 $\pm$ 0.271  \\
J1307+5427      &         -60 $\pm$           8  &        -275 $\pm$          17 &15.71 $\pm$ 0.05& 3.87 $\pm$ 1.01  & 0.680 $\pm$ 0.224  \\
J1315+6207      &         -97 $\pm$           5  &        -481 $\pm$          32 &15.34 $\pm$ 0.31& 4.53 $\pm$ 2.31  & 0.110 $\pm$ 0.060  \\
J1403+0628      &        -136 $\pm$          19  &        -416 $\pm$          64 &15.16 $\pm$ 0.53& 2.93 $\pm$ 2.48  & 0.138 $\pm$ 0.120  \\
GP1054+5238     &        -160 $\pm$          16  &        -409 $\pm$          61 &15.42 $\pm$ 0.08& 9.52 $\pm$ 2.43  & 0.206 $\pm$ 0.067  \\
\enddata
\label{tab:outflow}
\tablecomments{Table of outflow properties, derived from the COS spectra. Column 1 gives the galay name. Column 2 gives the centroid velocity (v$_\text{cen}$) of the Si~{\sc II} absorption line, while column 3 gives the velocity at 90\% of the continuum (\vnp) of the Si~{\sc II} absorption line. The logarithm of the column density of the Si~{\sc II} absorption line is given in column 4. Column 5 gives the mass outflow rate of the warm (roughly 10$^4$~K) gas. Column 6 gives the mass loading factor ($\eta = $M$_\text{o}$/SFR) of the warm gas.}
\end{deluxetable}
\pagebreak
\clearpage
\bibliography{hstphases_arxiv}

\end{document}